\documentclass{article}[11pt]
\usepackage{jheppub}
\usepackage[toc,page]{appendix}
\pdfoutput=1
\usepackage{graphicx}
\usepackage{amsmath,amssymb,mathrsfs}
\usepackage{bbm}
\usepackage{color}
\usepackage{xcolor}
\usepackage{dsfont}
\usepackage{cancel}
\usepackage{dsfont}
\usepackage{epstopdf}
\usepackage{epsfig}
\usepackage{bm}
\usepackage{dcolumn}
\usepackage{hyperref}
\usepackage{enumitem}
\usepackage{multirow}
\usepackage{lineno}
\usepackage{mathtools,slashed}
\usepackage{url}
\usepackage{lineno}
\usepackage{wrapfig}
\usepackage[load-configurations=abbreviations, redefine-symbols=true]{siunitx}

\newcommand{\ben}{\begin{enumerate}}
\newcommand{\een}{\end{enumerate}}
\newcommand{\bit}{\begin{itemize}}
\newcommand{\eit}{\end{itemize}}

\newcommand{\beqa}{\begin{eqnarray}}
\newcommand{\eeqa}{\end{eqnarray}}
\newcommand{\beq}{\begin{equation}}
\newcommand{\eeq}{\end{equation}}
\newcommand{\bay}{\begin{array}}
\newcommand{\eay}{\end{array}}

\def\ifmath#1{\relax\ifmmode #1\else $#1$\fi}

\def\gsim{\ \rlap{\raise 3pt \hbox{$>$}}{\lower 3pt \hbox{$\sim$}}\ }
\def\lsim{\ \rlap{\raise 3pt \hbox{$<$}}{\lower 3pt \hbox{$\sim$}}\ }

\def\ls#1{\ifmath{_{\lower1.5pt\hbox{$\scriptstyle #1$}}}}
\def\lsup#1{^{\lower 6pt\hbox{$\scriptstyle#1$}}}

\def\bracket#1#2 {\mathinner{\langle{#1}|{#2}\rangle}}

\def\bracket#1#2 {\mathinner{\langle{#1}|{#2}\rangle}}

\newcommand{\be}{\begin{equation}}
\newcommand{\ee}{\end{equation}}
\newcommand{\bea}{\begin{eqnarray}}
\newcommand{\eea}{\end{eqnarray}}

\graphicspath{{figs/}}

\begin{document}

\title{AION: An Atom Interferometer Observatory and Network}

\abstract{
We outline the experimental concept and key scientific capabilities of AION (Atom Interferometer Observatory and Network), a proposed experimental programme using cold strontium atoms 
to search for ultra-light dark matter, to explore gravitational waves in the mid-frequency 
range between the peak sensitivities of the LISA and LIGO/Virgo/ KAGRA/INDIGO/Einstein Telescope/Cosmic Explorer
experiments, and to probe other frontiers in fundamental physics. AION would complement other planned
searches for dark matter, as well as probe mergers involving intermediate-mass black holes and explore early-universe cosmology. AION would share many technical features with the MAGIS experimental
programme, and synergies would flow from operating AION in a network with this experiment, as well as with other atom interferometer experiments such as MIGA, ZAIGA and ELGAR. Operating AION in a network with other gravitational wave detectors such as LIGO, Virgo and LISA would also offer many 
synergies. \\
~~\\
~~\\
~~\\
{AION-2019-001}
}


\author[1]{L.~Badurina,}
\affiliation[1]{Department of Physics, King's College London, Strand, London, WC2R 2LS, UK}


\author[2]{E.~Bentine,}
\affiliation[2]{Department of Physics, University of Oxford, Keble Road, Oxford, OX1 2JJ, UK}

\author[1]{D.~Blas,}

\author[3]{K.~Bongs,}
\affiliation[3]{MUARC, Physics and Astronomy, University of Birmingham, Edgbaston, Birmingham, B15 2TT, UK}

\author[2]{D.~Bortoletto,}

\author[4]{T.~Bowcock,}
\affiliation[4]{Department of Physics, University of Liverpool, Merseyside, L69 7ZE, UK}
 
\author[4]{K.~Bridges,}

\author[5,6]{W.~Bowden,}
\affiliation[5]{National Physical Laboratory, Teddington, Middlesex, TW11 0LW, UK}
\affiliation[6]{Department of Physics, Blackett Laboratory, Imperial College, Prince Consort Road, London, SW7 2AZ, UK}

\author[6]{O.~Buchmueller,}
\emailAdd{ o.buchmueller@imperial.ac.uk}

\author[7]{C.~Burrage,}
\affiliation[7]{School of Physics and Astronomy, University of Nottingham, Nottingham, NG7 2RD, UK}

\author[4]{J.~Coleman,}

\author[4]{G.~Elertas,}

\author[1,8,9]{J.~Ellis,}
\affiliation[8]{National Institute of Chemical Physics \& Biophysics, R{\"a}vala 10, 10143 Tallinn, Estonia}
\affiliation[9]{Theoretical Physics Department, CERN, CH-1211 Geneva 23, Switzerland}

\author[2]{C.~Foot,}

\author[10]{V.~Gibson,}
\affiliation[10]{Cavendish Laboratory, J J Thomson Avenue, University of Cambridge, CB3 0HE, UK}


\author[11, 12]{M.~G.~Haehnelt,}
\affiliation[11]{Kavli Institute for Cosmology, Madingley Road, Cambridge, CB3 0HA, UK}
\affiliation[12]{Institute of Astronomy, Madingley Road, Cambridge, CB3 0HA, UK}

\author[10]{T.~Harte,}

\author[3]{S.~Hedges,}

\author[5,6]{R.~Hobson,}

\author[3]{M.~Holynski,}

\author[4]{T.~Jones,}
 
\author[3]{M.~Langlois,}
\author[3]{S.~Lellouch ,}
 
\author[1,13]{M.~Lewicki,}
\affiliation[13]{Faculty of Physics, University of Warsaw, ul. Pasteura 5, 02-093 Warsaw, Poland}

\author[10,11]{R.~Maiolino,}

\author[14]{P.~Majewski,}
\affiliation[14]{STFC Rutherford Appleton Laboratory (RAL), Didcot, OX11 0QX, UK}
\author[6]{S.~Malik,}
\author[15]{J.~March-Russell,}
\affiliation[15]{Rudolf Peierls Centre for Theoretical Physics, Oxford University, OX1 3PU, UK}

\author[1]{C.~McCabe,}

\author[14]{D.~Newbold,}

\author[6]{B.~Sauer,}

\author[10]{U.~Schneider,}

\author[2]{I.~Shipsey,}

\author[3]{Y.~Singh,}


\author[10]{M. A.~Uchida,}

\author[14]{T.~Valenzuela,}

\author[14]{M.~van~der~Grinten,}

\author[1,8]{V.~Vaskonen,}

\author[4]{J.~ Vossebeld,}

\author[2]{D.~Weatherill,}

\author[14]{I.~Wilmut}

\maketitle


\newpage
\section{Introduction}

This article describes a proposal to construct and operate a next-generation Atomic Interferometric Observatory and Network (AION) that will enable the exploration of properties of dark matter (DM) and searches for new fundamental interactions. In addition, it will provide a pathway towards detecting gravitational waves (GWs) from the very early Universe and astrophysical sources in the mid-frequency band ranging from $\sim 0.01$~Hz to a few Hz, where currently operating and planned detectors are relatively insensitive. 

We outline a staged plan to build a set of atom interferometers with baselines increasing from an initial vertical length of 10m, which will pave the way for 100m and eventually km-scale terrestrial detectors,  and ultimately a satellite-based detector in the future.  AION will use quantum sensors that are based on the superposition of atomic states, like the MAGIS programme~\cite{Graham:2017pmn,Coleman:2018ozp}. These devices combine established techniques from inertial sensing with new features used by the world's best atomic clocks~\cite{Snadden:1998zz,Dimopoulos:2008sv,Graham:2012sy,Graham:2016plp,Graham:2017pmn}. The AION programme will enable cutting-edge exploitation of the enormous physics potential of the mid-frequency band, with several opportunities for ground-breaking discoveries. It will also develop the foundation for future science with ultra-sensitive quantum sensors and for a new and potentially highly-disruptive class of applications of precision measurement in surveying and prospecting.

The full AION programme consists of 4 stages. Stage 1 will build and commission the 10m detector, produce detailed plans and predictions for the performance of a 100m device, and develop technologies to meet the requirements of the 100m device. Stage 2, which is not part of the initial proposal, will build, commission and exploit the 100m detector and prepare a design study for the kilometre-scale terrestrial detector. Stages 3 and 4, a km-scale and satellite-based (thousands of kilometres scale) detectors, respectively~\cite{Bertoldi:2019tck}, are the long-term objectives of the continuing programme.

Through its unparalleled sensitivity to the physics of space-time and its distortion between the sensors, AION's science case has broad applications to fundamental physics and aligns well with the highest priorities of several international science communities. 

Firstly, the detector is highly sensitive to time-varying signals that could be caused by ultra-light bosons. The discovery of such particles and their associated fields could reveal the nature of DM and blueprint a novel method to probe the associated theoretical frameworks. There are several such candidate ultra-light bosons - including dilatons, relaxions, moduli, axions and vector bosons - that are able to produce a signal in the frequency range 100 nHz to 10 Hz. Such hidden-sector particles could play a crucial role in particle physics beyond the Standard Model (SM), astrophysics, and cosmology.
The early stages of AION already have the potential to search for ultra-light dark matter candidates in a large mass range from $\sim 10^{-12}$ to $\sim 10^{-17}$ eV with unprecedented sensitivity, and networking with MAGIS offers interesting ways to characterize a signal.

Secondly, AION is sensitive to GWs within a frequency range that lies between the bands where the LIGO~\cite{TheLIGOScientific:2014jea},  Virgo~\cite{TheVirgo:2014hva} and LISA~\cite{Audley:2017drz} experiments are sensitive, thus opening a new window on the cosmos~\cite{Dimopoulos:2006nk,Dimopoulos:2007cj}. It is expected that the GW spectrum from $\sim 0.01$~Hz to $\sim 10$~Hz will be mostly free of continuum foreground noise from astrophysical sources such as white dwarves. Possible sources in this frequency range include mergers of intermediate-mass black holes, first-order phase transitions in the early Universe and cosmic strings. Moreover, there are interesting prospects for synergies via networking with measurements by MAGIS~\cite{Graham:2017pmn,Coleman:2018ozp} in a similar wavelength band, and with LISA and LIGO measurements in complementary wavelength bands.

Thirdly, AION may also be able to explore other aspects of fundamental physics such as fifth forces, the equivalence principle, variations in fundamental constants, dark energy and other basic physical principles.

The AION programme may be summarized as follows:
\begin{itemize}
\item To build a series of instruments exploiting recent advances in cold atoms to explore fundamental issues in physics, astrophysics and cosmology, including;
\item A new generation of generic precision searches for new ultra-light particles and their fields complementing those performed at collider facilities; and
\item To explore mid-frequency band GWs from astrophysical sources and the very early Universe, such as space-time `tremors' produced by astrophysical sources, phase transitions in the early universe and cosmic strings.
\end{itemize}

%
%

This last theme offers long-term outputs of high scientific value, is a cross-over between traditional particle physics, astrophysics and the physics of the early Universe, and opens new prospects for multi-messenger observations. In particular, with the 100m stage and the eventual construction of the km-scale atom interferometer detector, new GW sources will become observable. AION will establish large-scale interferometric infrastructure and techniques and
complement the observational breadth of LISA, LIGO  and other operating and approved detectors, crucially allowing complete coverage of the frequency spectrum and offering interesting networking opportunities. The 100m baseline atom interferometer would, on its own, be sensitive to mergers of ${\cal O}(10^4)$ solar-mass black holes. At the 1 km scale, sources such as neutron star binaries or black hole mergers will be observed in the mid-frequency band, prior to later observations by LIGO, Virgo, KAGRA~\cite{Somiya:2011np} and INDIGO~\cite{Unnikrishnan:2013qwa} after the merger has passed to higher frequencies. AION observations will be a powerful new source of information, giving a prediction of the time and location of a merger event in those detectors – potentially months before it occurs. Likewise, early inspiral stages of mergers may be observed at lower frequencies by LISA, leading later to mergers measured by AION.

From the outset, AION will benefit strongly from close collaboration on an international level with MAGIS-100~\cite{Graham:2017pmn}, which pursues a similar goal of an eventual km-scale atom interferometer. The AION programme would reach its ultimate sensitivity by networking two detectors simultaneously,
providing unique physics opportunities not accessible to either detector alone.
Collaboration with AION has already been endorsed by the MAGIS Collaboration.
In addition to being a vital ingredient of our short- and mid-term objectives, this collaboration will serve as the test bed for full-scale terrestrial (kilometre-scale) and satellite-based (thousands of kilometres scale) atom interferometers and build the framework for global scientific collaboration in this area, which could include other groups (e.g., MIGA~\cite{Canuel:2017rrp}, ELGAR~\cite{Canuel:2019abg} and ZAIGA~\cite{Zhan:2019quq}). 

The layout of this article is as follows. In Section~\ref{concept} we describe the experimental concept of AION, which is based on an interferometer using strontium atoms, and outline the basic parameters and performance and goals of successive stages of the AION programme. In Section~\ref{ULDM} we discuss the prospective sensitivities of the successive AION stages for detecting ultra-light scalar and vector dark bosons. The latter offers interesting discovery possibilities for AION-10, and AION-100 also has interesting capabilities for detecting ultra-light scalar bosons. We discuss in Section~\ref{GW} the prospects for detecting and measuring GWs from astrophysical sources, particularly mergers of intermediate-mass black holes, and particle processes in the early universe such as first-order phase transitions and cosmic strings, casting light on the early evolution of the universe. The opportunities offered by networking with MAGIS, LIGO and Virgo are stressed in Subsection~\ref{network}. Opportunities to explore aspects of fundamental physics such as the equivalence principle are discussed in Section~\ref{FP}, and Section~\ref{FR} contains some final remarks.

\section{Experimental Concept}
\label{concept}

The  experimental  concept for the AION project is similar to that proposed in \cite{Dimopoulos:2008sv,Graham:2012sy,Graham:2016plp,Graham:2017pmn} and is based on a differential phase measurement between several atom interferometers (AIs) using atoms in free-fall that are operated simultaneously using a common laser source \cite{Snadden:1998zz}. 
In the following we outline a schematic design of the experimental concept, while pointing out that several of its details are yet to be defined. The interferometry is performed in a vertical vacuum system of length $L$, which is shielded from residual magnetic fields. Atom sources that prepare and launch the cold atomic clouds used in the AI sequence are positioned along the length of the vacuum system.  As discussed in more detail below, ultralight dark matter (ULDM) is detected via its differential effects on the atomic transition frequencies, while
a passing gravitational wave (GW) is detected via the strain it creates in the space between the free-falling atoms. 
This strain changes the light propagation time, and therefore the difference in the laser phase measured by the spatially-separated AIs. 

\begin{figure}[t]
\centering
\includegraphics[width=0.4\textwidth]{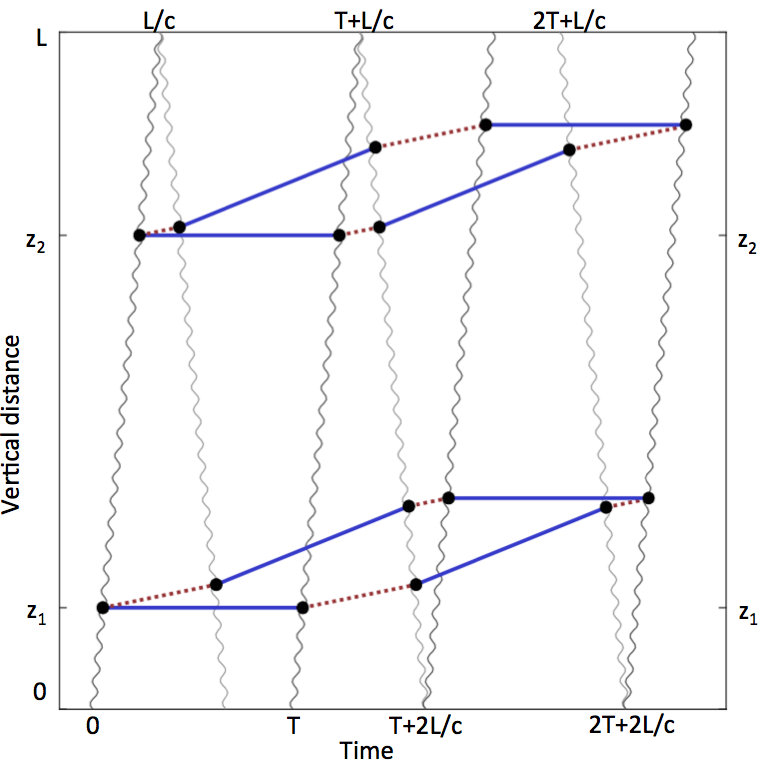}
\caption{\it
Space-time diagram of the operation of a pair of cold-atom interferometers based on single-photon transitions between the ground state (blue) and the excited state (red dashed). Height in the AION configuration is shown on the vertical axis and the time axis is horizontal.
The laser pulses (wavy lines) travelling across the baseline from opposite sides are used to divide, redirect, and recombine the atomic de Broglie waves, yielding interference patterns that are sensitive to the modulation of the atomic transition frequency caused by coupling to DM, or the modulation of the light travel time caused by 
GWs. For clarity, the sizes of the atom interferometers are shown on an exaggerated scale.}
\label{space-time}
\end{figure}

The principle of operation is illustrated in Fig.~\ref{space-time} for the case of two vertically-separated AIs. The interferometers are located at $z_1$, $z_2$, separated by a large vertical distance $L \approx z_2 - z_1$, and are represented by the two diamond-shaped loops on an enlarged scale. 
The laser pulses (wavy lines) drive transitions between the ground and excited states of the atoms and in the process transfer momenta $\hbar k$ to the atoms. Here, $k=2\pi/\lambda$ denotes the wave-vector of the photons. 
By engineering an appropriate sequence of pulses from alternating directions, the laser pulses act as beam splitters and mirrors for the atomic de Broglie waves, generating a quantum superposition of two paths and then recombining them. The phase imprinted on each arm of the AI depends both on the phase of the laser pulses that excite and de-excite the atoms, and on the phase accumulated by the atoms themselves due to energy shifts. The final pulse sequence  acts as the last beam splitter and superimposes the matter-waves from the two interferometer arms, generating interference. The fraction of atoms leaving in each exit port, i.e.\ in the ground or excited state, is readily detected and provides a direct measure of the relative phases accumulated along each interferometer arm. While the resulting phase depends critically on the overall laser phase, there is a crucial advantage to using several interferometers within the same vacuum tube: the same laser beam can be used in several interferometers and the effect of laser phase noise can therefore be  suppressed efficiently by making differential phase measurements $\Delta \phi_{\rm diff}$ between e.g.\ the two interferometers at $z_1$ and $z_2$.

The presence of a gravitational wave of strain amplitude~$h$ and frequency~$\omega$ modulates the AI separation distance $L$, giving rise to time variations in~$\Delta \phi_{\rm diff}$. A similar signal might also be produced by ULDM, which could induce a small time-dependent perturbation to the atomic transition frequency~$\omega_A$ as the dark matter field evolves~\cite{Arvanitaki:2016fyj}. Since the laser interacts with the separate AIs at different times due to the light propagation delay, a DM-induced time-dependent perturbation will be observable as fluctuations in the differential phases accumulated by the separate AIs.

Using AIs to detect GWs was initially proposed for Raman-based AIs, which use two-photon transitions to perform the required atom-optic pulses~\cite{Dimopoulos:2008hx}.
However, the phase fluctuations of the laser fields used to drive the two-photon transitions introduce additional noise, because wave packets interact with light fields emitted from the lasers at different times.
This noise grows with $L$, making long-baseline interferometry based on two-photon transitions unfeasible.
Instead, this measurement scheme leverages recent advances in optical frequency control to realise AIs operating on a single-photon optical transition between a ground state and 
long-lived `clock' state, separated in energy by~$\hbar\omega_A$~\cite{Hu:2017kp}.
Using single-photon transitions allows all AIs to be addressed with a pulse of light emitted from the laser source at the same instant, thereby suppressing the effect of phase fluctuations of the driving laser field.

The metric (in TT gauge) for a plane gravitational wave traveling in the $x$-direction  is~\cite{Graham:2012sy}
\begin{equation}
d s^{2}=c^2 d t^{2}-\left(1+h \sin \left(\omega(t-z)+\phi_{0}\right)\right) d z^{2}-\left(1-h \sin \left(\omega(t-z)+\phi_{0}\right)\right) d y^{2}-d x^{2} \;,
\label{metricGW}
\end{equation}
where $\phi_0$ is the phase of the gravitational wave at the start of AI pulse sequence. Ref.~\cite{Hoganthesis} has shown that the leading contribution to the phase shift for a single-photon transition in a single interferometer located at $z_1$ is
\begin{equation}
4 \frac{h \,\omega_{A}}{\omega} \sin ^{2}\left(\frac{\omega T}{2}\right) \sin \left(\frac{\omega z_{1}}{2c}\right) \sin \left(\omega T+\frac{\omega z_{1}}{2c}+\phi_{0}\right)\,,\\
\label{leading}
\end{equation}
where $2T$ is the interrogation time. 

When simultaneously operating \textit{two} atom interferometers located at \(z_1\) and \(z_2\) 
the leading order contribution to the differential phase shift arises from the difference between Eq.~\eqref{leading} evaluated at \(z_1\) and~\(z_2\), namely:
\begin{equation}
\Delta \phi_{\rm{diff}} = 4 \frac{h\, \omega_{A}}{\omega} \sin^{2} \left(\frac{\omega T}{2}\right) \sin \left(\frac{\omega \left(z_{2}-z_{1}\right)}{2 c} \right) \sin \left(\omega T+\frac{\omega \left(z_{1} + z_{2} \right)}{2c} + \phi_{0}\right)  \;,
\end{equation}
which in the limit \(z_2-z_1 \approx L\) and \(\omega L/c \ll 1 \), is
\begin{equation}
\Delta \phi_{\rm{diff}} \approx 2 h k L \sin^{2} \left(\frac{\omega T}{2}\right) \sin \left ( \omega T + \phi_0 \right)\;,
\label{single photon}
\end{equation}  
where $k \approx \omega_A/c$ when operating with single-photon transitions.
For the AION project, we intend to operate the AI using strontium, as it has both convenient transitions for laser cooling and a long-lived metastable state which can be addressed using an ultra-stable (clock) laser at \SI{698}{\nano\meter}.

We intend to employ large momentum transfer (LMT) techniques to amplify the phase shift imparted by the laser field to the atoms. For instance, after the initial beam splitter pulse, $n-1$ $\pi$-pulses can be applied from opposite directions, exciting and de-exciting the atoms. Each pulse imparts the phase of the laser onto the atomic wave function, along with an increase in momentum of $\hbar k$, where $k = \omega_A/c$.
This enhances the differential gravitational wave signal by a factor of $n$. 
We note in this connection that strontium-based single-photon AI has recently demonstrated 141 $\hbar k$ LMT~\cite{Rudolph:2019vcv}, 
and comment also that pulse-shaping techniques will also be used to improve the fidelity of LMT sequences.

For higher-frequency GWs that oscillate several times during the AI interrogation time $2T$, the effect of the GW would largely cancel in the above configuration. This reduction in signal can however be mitigated by operating the AI in a resonant mode. As outlined in details  in~\cite{Graham:2016plp}, this can be accomplished by using the pulse sequence $\pi/2 - \pi -  \dots - \pi - \pi/2$  with $Q$ $\pi$-pulses instead of the standard, broadband $\pi/2 - \pi - \pi/2$ pulse sequence. 

\begin{wrapfigure}[27]{lt}{0.43\textwidth}
\vspace{-0.cm}
\includegraphics[width=0.43\textwidth]{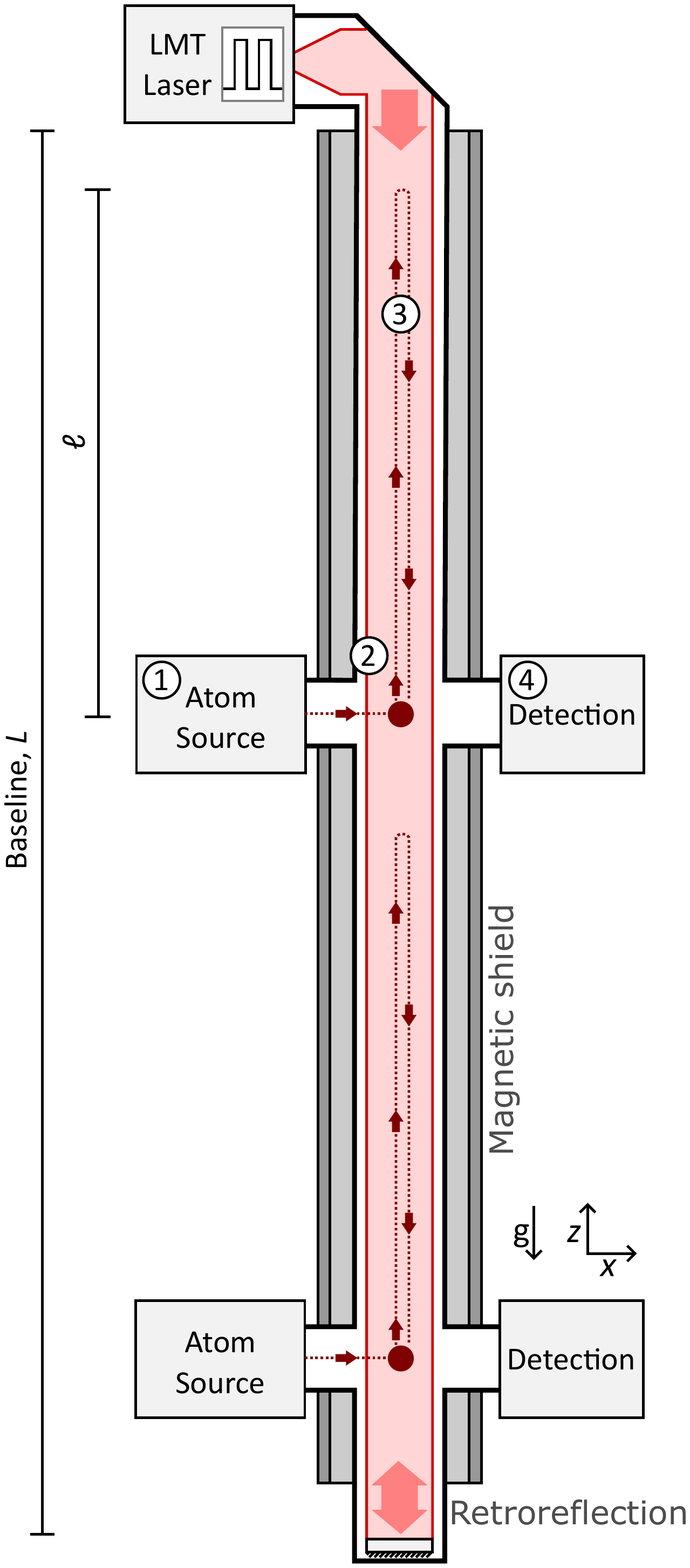}
\vspace{-0.5cm}
\caption{\it Conceptual scheme of AION, illustrated for two atom interferometers that are arranged vertically and addressed by a single laser source.}
\label{Schematic}
\end{wrapfigure}
Fig.~\ref{Schematic} illustrates the conceptual scheme of AION, for two AIs addressed by a single laser source and arranged vertically. 
Each AI contains a source of ultra-cold atoms~(1), which are transported to the centre of the vacuum system and launched vertically~(2) (launch optics are not shown).
The clouds are in free fall for a time $T_\text{fall}=\sqrt{8\ell/g}$~(3),

during which the atom interferometer sequence is performed using light from the clock laser that simultaneously addresses all the atom interferometers.
Finally, the phase accumulated by each atom interferometer is read out individually by imaging the atom clouds~(4).
Grey boxes indicate the subsystems responsible for producing laser light at the clock transition, the sources of ultra-cold atoms, and the detection optics and readout.
The path taken by the atomic clouds is indicated by the dark red dashed line, and an offset has been added between the upward and downward travelling directions for clarity.
The extent of the vacuum system is shown by the thick black line, and the surrounding magnetic shield is shown in grey.
Vacuum pumps etc.\ are omitted  for clarity.

For two atom interferometers operating in resonant mode, the phase response of the detector can be written as $\Delta\Phi_\text{grad}(t_0)=\Delta\phi \cos{(\omega t_0+\phi_0)}$, where $\omega t_0+\phi_0$ is the phase of the GW at time $t_0$, which denotes the start of the pulse sequence. The amplitude of the detector response is derived in~\cite{Graham:2016plp}, and is given~by:
\begin{equation}
\Delta\phi= n k h L \frac{\sin\!{(\omega Q T)} }{\cos\!{(\omega T/2)}}\,\mathrm{sinc}{\left(\frac{\omega n L}{2 c}\right)}  \,\sin\!{\left(\frac{\omega T}{2}\!-\!\frac{\omega(n-1)L}{2c}\right)} \, ,
\label{Eq:PhaseShift1}
\end{equation}
which is peaked at the resonance frequency 
$\omega_r \equiv \pi/T$ and has a bandwidth $\sim\!\omega_r/Q$. On resonance the amplitude of the peak phase shift is:
\begin{equation}
\qquad \Delta\phi_\text{res}=2Q n k h L\,\mathrm{sinc}{\left(\frac{\omega_r n L}{2 c}\right)} \,\cos\!{\left(\frac{\omega_r (n-1) L}{2 c}\right)} \, ,
\label{Eq:PeakPhaseShift}
\end{equation}
which reduces to $\Delta\phi_\text{res}\approx 2 Q n k h L$ in the low-frequency limit $\omega_r \ll \tfrac{c}{n L}$. Large momentum transfer operation enhances $n$-fold the sensitivity of the phase response. By changing the pulse sequence used to operate the device (changing $Q$)~\cite{Graham:2016plp} the interferometer can be switched from broadband to resonant mode, resulting in a $Q$-fold enhancement.

We assume in making the AION sensitivity projections for GW signals that the detector is operated mainly in the resonant mode, while for DM signals we provide projections for both resonant and broadband modes. In order to generate the sensitivity curve for, e.g., a GW signal, from the phase response, we calculate the minimum strain $h$ that is detectable given a phase noise spectral density $\delta\phi_{\rm noise}$. We optimize the LMT enhancement $n$ for each frequency and resonant enhancement $Q$, taking into account the detector design constraints, which include the limits on the total number of pulses, $n_p^{\rm max} = 2Q(2n-1)+1$, and on the maximum interferometer duration, $2TQ < T_{\rm max}$, where $T_{\rm max}$ is the maximum time over which the atom interferometer is interrogated.

 This resonant mode strategy provides significant sensitivity to a stochastic background of GWs,
 {  e.g., of cosmological origin. To indicate the sensitivity estimates for the density of GW energy, $\Omega_\text{GW}$, we use power-law  integration~\cite{Thrane:2013oya} to display an envelope of power-law signals for each given frequency detectable with an assumed signal-to-noise ratio ${\rm SNR}=10\, $.  In the calculations for the different stages of AION we assume goals of five years of observation time divided between 10 logarithmically-distributed resonance frequencies and sum the signal from the total running time of the experiment.}
  {We have verified that changing this scanning strategy by using a different number of resonant frequencies does not have a strong impact on the resulting sensitivity.}
  {These curves thus have the property that any power-law signal touching them would give the required SNR in the indicated experiment. For ease of comparison, we also assumed five years of operation for each of the other experiments shown.} 
  
 We adopt the following projection scenarios for the various stages of AION:
\begin{itemize}

\item AION-10-initial: This scenario represents the sensitivity estimate of a \SI{10}{\meter} detector using basic parameter estimates that are achievable today. This will be the reference to benchmark future improvements.
\item AION-10-goal: This scenario represents the sensitivity estimate of a \SI{10}{\meter} detector using parameters that we plan to achieve as goals for the 10m stage. This sets the benchmark for the ultimate sensitivity of the 10m detector and is also the starting point for AION-100. 

\item AION-100-initial: This scenario represents the sensitivity estimate of a \SI{100}{\meter} detector using basic parameter estimates that are achievable at the time AION-100 will start operation. This will be the reference to benchmark future improvements for AION-100.
\item AION-100-goal:  This scenario represents the sensitivity estimate of a \SI{100}{\meter} detector using parameters that we plan to achieve as goals for the \SI{100}{\meter} stage. This sets the benchmark for the ultimate sensitivity of the  \SI{100}{\meter} detector.
\item AION-km:  This scenario represents the sensitivity estimate of a km-scale detector using parameters that we plan to achieve as goals for this stage of the project. This sets the benchmark for the sensitivity of the km-scale detector.

\end{itemize}
The values of the basic parameters assumed for the different sensitivity scenarios are listed in Table~\ref{tab:parameters}. These basic parameters mainly determine the sensitivities of the projections. To benchmark the sensitivity of a space-based detector, we use the results from the proposed  AEDGE~\cite{Bertoldi:2019tck} two-satellite mission, which is based on the same concept as the AION programme.

\begin{table}[t!]
  \begin{center}
    \begin{tabular}{c|c|c|c|c} 
    \textbf{Sensitivity} & $L$ & $T$ & $\delta\phi_{\rm noise}$ & LMT\\
      \textbf{Scenario} & [m] & [sec] & [$1/\sqrt{{\rm Hz}}$] & [number $n$]\\
      \hline
      AION-10 (initial) & 10 & 1.4 & $10^{-3}$ & 100\\
      AION-10 (goal) & 10 & 1.4 & $10^{-4}$ & 1000\\
      AION-100 (initial) & 100 & 1.4 & $10^{-4}$ & 1000\\
      AION-100 (goal) & 100 & 1.4  & $10^{-5}$ & 40000\\
      AION-km & 2000 & 5 & $0.3 \times 10^{-5}$ & 40000\\
    \end{tabular}
      \caption{\it List of basic parameters: length of the detector L; interrogation time of the atom interferometer~$2T$; phase noise $\delta\phi_{\rm noise}$; and number $n$ of large momentum transfers (LMT). The choices of these parameters largely determine the sensitivities of the projection scenarios. 
    In our projection scenarios for AION-10 and AION-100, we assume $T=1.4$~\!s, which corresponds to a 10~\!m atom interferometer setup in launch mode. The exact value for $T$ will depend on the precise design and setup for the different stages, which have yet to be finalised.
    }\label{tab:parameters}
  \end{center}
\end{table}

\section{Ultra-Light Dark Matter}
\label{ULDM}

The cosmological standard model successfully describes physics over a large range of scales. 
It requires the existence of DM, an invisible form of matter that provides around 84\% of the matter density in the Universe~\cite{Aghanim:2018eyx}. 
All the available evidence for DM is due to its gravitational interaction, but it is widely expected that DM interacts with conventional matter also through interactions other than gravity, though any such interactions of DM particles must be very weak. 
Direct searches for DM particles via  
their non-gravitational interactions with terrestrial matter are among the highest priorities in particle physics, astrophysics and cosmology. 
There have been many direct searches for weakly-interacting massive particles (WIMPs), whose mass lies in the GeV to multi-TeV window, and current experiments probe interaction cross-sections well below
the weak scale. As yet, there are no positive results from such experiments (see, e.g., the constraints from the XENON1T experiment~\cite{Aprile:2018dbl}), nor from searches for WIMP production at the LHC or indirect searches for annihilations of astrophysical WIMPs.

Although theoretical extensions of the SM of particle physics provide many WIMP DM candidates, they also offer elementary particle DM candidates over a much broader mass range, from $10^{-22}$ eV to the Planck scale ($\sim10^{18}$~GeV)~\cite{Battaglieri:2017aum}. 
Among the most interesting alternatives to WIMPs are
ultra-light bosons with a sub-eV mass, which owing to their large occupation number, are expected to be behave as coherent waves. Prominent among these ultra-light scalar bosons are many well-motivated DM candidates (for reviews, see~\cite{Jaeckel:2010ni,Battaglieri:2017aum}), such as dilatons, moduli and the relaxion~\cite{Graham:2015cka}, as well as the pseudoscalar QCD axion and axion-like-particles (ALPs)~\cite{Marsh:2017hbv}, and (dark) vector bosons~\cite{Graham:2015ifn}. These ultra-light bosons are also viable DM candidates, as they may well have acquired the observed abundance, e.g., through the misalignment mechanism~\cite{Preskill:1982cy,Abbott:1982af, Dine:1982ah}  or quantum fluctuations during inflation~\cite{Graham:2018jyp,Guth:2018hsa,Ho:2019ayl}, and such DM bosons are naturally cold, as required by the standard cosmological paradigm for astrophysical structure formation. Direct searches for ultra-light bosonic DM particles are challenging, but are now assuming enhanced importance.

\subsection{Scalar dark matter}

Atom interferometers can detect a distinctive prediction of ultra-light scalar DM~\cite{Geraci:2016fva, Arvanitaki:2016fyj}. 
Interactions of ultra-light bosons with electrons and photons cause oscillations in the
electron mass and electromagnetic fine-structure constant, with an amplitude set by the local DM density and a frequency given by the mass of the scalar DM particle. These oscillations cause, in turn, oscillations in atomic transition frequencies, which depend on the electron mass and fine-structure constant. 

A differential atom interferometer will measure a non-trivial signal phase when the period of the DM wave oscillation matches the total duration of the interferometric sequence. Since all phases in the interferometric sequence cancel in the absence of matter-DM interactions, only the DM-induced phase $\Phi^{t_1}_{t_0}$ accumulated in the excited state between times $t_0$ and $t_1$, relative to the phase of the ground state, contributes to the dark matter signal~\cite{Arvanitaki:2016fyj}. 
In AION,
the signal channel corresponds to the difference between the total accumulated phase in each interferometer and takes the form
\begin{equation}
\hspace{-0.5cm}
\Phi \simeq \Phi^{T+L/c}_{T-(n-1)L/c} - \Phi^{nL/c}_{0} - \left[\Phi^{2T+L/c}_{2T-(n-1)L/c} - \Phi^{T+nL/c}_{T}\right] \, ,
\end{equation}
where, as defined previously, $L$ is the distance between the two atom interferometers, $n$ is the number of large-momentum transfer kicks and $c$ is the speed of light.

The simplest possibility is that the ultra-light scalar DM couples {\it linearly} to Standard Model fields~\cite{Damour:2010rm,Damour:2010rp} through some interaction of the form
\begin{equation}\label{linear}
\mathcal{L}^{\rm{lin}}_{\rm{int}}  \supset - \phi(\mathbf{x},t) \cdot \sqrt{4 \pi G_{\rm{N}}} \cdot \left[ d_{m_e} m_e \bar{e}e - \frac{1}{4} d_e F_{\mu \nu} F^{\mu \nu} \right]  + b\, \phi(\mathbf{x},t) |H|^2    \;,
\end{equation}
where $G_{\rm{N}}$ is Newton's constant, $m_e$ is the electron mass, $d_{m_e}$, $d_e$ and $b$ are the couplings of scalar DM to normal matter and we have used units where $\hbar=c=1$.
The large DM occupation number means that the scalar DM field behaves as a non-relativistic oscillating field approximated by
\begin{equation}
\phi(\mathbf{x},t)=\frac{\sqrt{2 \rho_{\rm{DM}}}}{m_{\phi}} \cos[m_{\phi}(t - \mathbf{v}_{\phi}\cdot \mathbf{x})+\cdots]\;,
\end{equation}
where $m_{\phi}$ is the scalar DM mass, $\rho_{\mathrm{DM}}$ is the time-and-space-dependent local DM density whose average value is $\sim 0.3~\mathrm{GeV}/\mathrm{cm}^3$, and $\mathbf{v}_{\phi}$ is the DM velocity whose magnitude $|\mathbf{v}_{\phi}|$ and dispersion $v_{\rm{vir}}$ have the characteristic value $\sim 10^{-3}c$. The ellipses in the argument of the cosine include 
an unknown random phase $\theta$ that encodes the coherence properties of the ULDM field: its coherence length is~\cite{Derevianko:2016vpm}
\begin{equation}
    \lambda_c = \frac{\hbar}{m_{\phi} v_{\rm{vir}}} \approx 2.0 \times 10^3 \,  \left( \frac{10^{-10}~\!{\rm eV} }{m_{\phi} c^2}  \right) \, {\rm km} \, , 
\end{equation}
and its coherence time is
\begin{equation}
 \tau_c = \frac{\hbar}{m_{\phi} v_{\rm{vir}}^2} \approx 6.6 \,  \left(  \frac{ 10^{-10}~\!{\rm eV}  }{m_{\phi} c^2 } \right)\, {\rm s} \, .
\end{equation}
This lack of perfect coherence must be taken into account when the integration time is less than the coherence time~\cite{Centers:2019dyn}. However, this is not a concern for the AION scenarios, since the assumed integration time ($10^8$~\!s) is always greater than the coherence time.
Owing to the small DM speed, we can also safely neglect terms $\mathcal{O}(|\mathbf{v}_{\phi}|^2)$.

For linear couplings, the signal amplitude (with $\hbar=c=1$) is~\cite{Arvanitaki:2016fyj}\begin{equation}
\bar{\Phi}_s = 8 \frac{\Delta \omega_A}{m_{\phi}} \left | \sin{\left[ \frac{m_{\phi}nL}{2} \right]}\sin{\left[ \frac{m_{\phi}(T-(n-1)L)}{2} \right]}\sin{\left[ \frac{m_{\phi}T}{2} \right]} \right | \, ,
\label{signal_amplitude}
\end{equation}
where
$\Delta \omega_A$ is the amplitude of the electronic transition oscillation induced by the scalar DM wave and can be expressed as 
\begin{equation}
\label{eq:deltaomega}
\Delta \omega_A = \frac{\sqrt{8 \pi G_{\rm{N}} \rho_{\rm{DM}}}}{m_{\phi}} \cdot  \omega_A \cdot \left(d_{m_e} + \xi_A d_e\right)  \, ,
\end{equation}
where $\xi_A$ is a  calculable parameter that depends on the chosen electronic transition. For the $5\mathrm{s}^2 \, ^1\mathrm{S}_0$ -- $5\mathrm{s}5\mathrm{p}\, ^3\mathrm{P}_0$ optical transition in strontium,  $\xi_A=2.06$~\cite{Angstmann:2004zz}.

\begin{figure}[t!]
\includegraphics[width=1.0\textwidth]{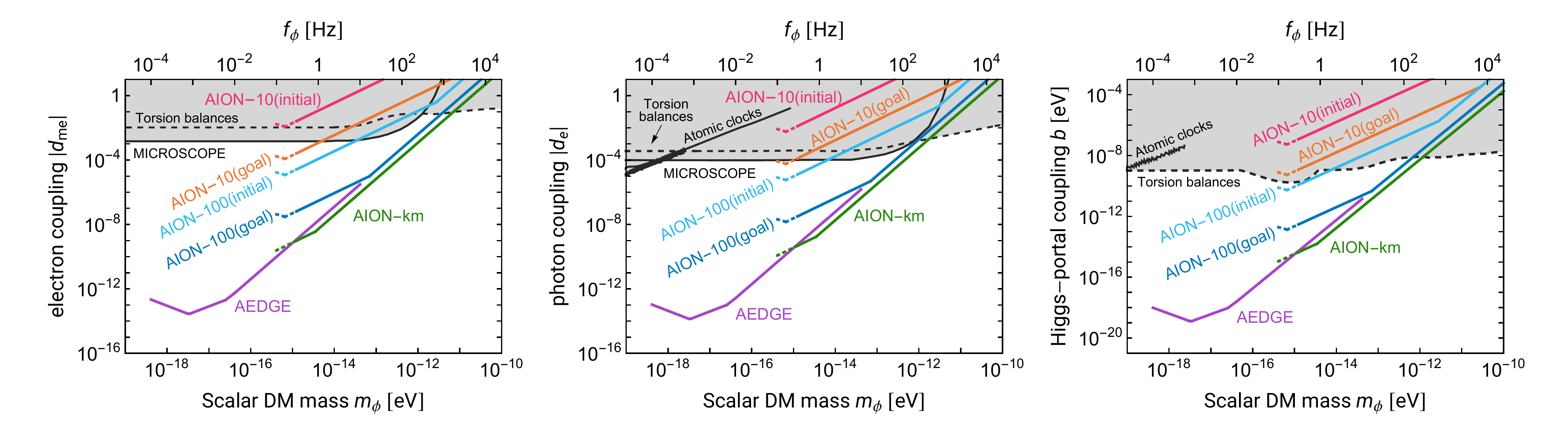}
\vspace{-0.35cm}
\caption{\it Sensitivities of different AION scenarios to scalar DM interactions with electrons (left), photons (middle) and the Higgs portal (right). The grey regions show parameter space that has already been excluded through searches for violations of the equivalence principle~\cite{Wagner:2012ui}, atomic spectroscopy~\cite{Hees:2016gop} by the MICROSCOPE experiment~\cite{Berge:2017ovy}.}
\label{DMplot}
\end{figure}

\begin{figure}[t!]
\includegraphics[width=1.0\textwidth]{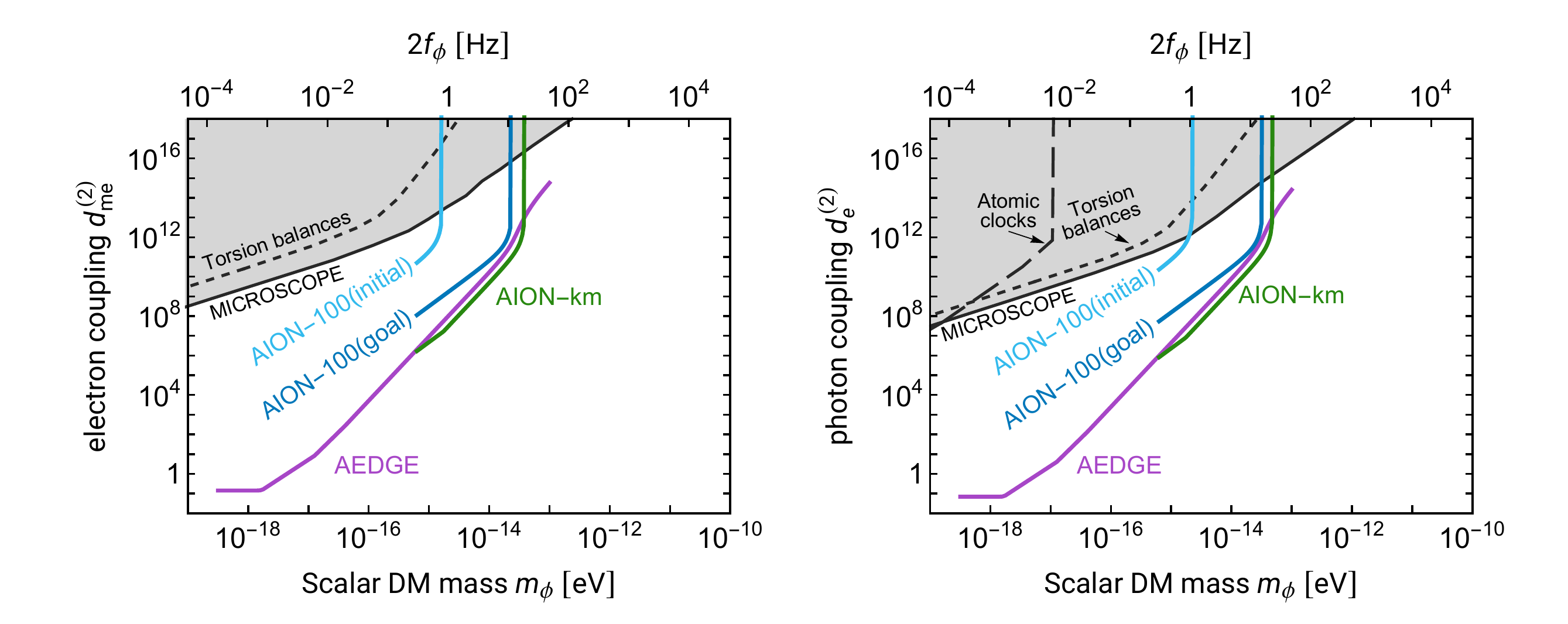}
\vspace{-0.5cm}
\caption{\it Sensitivities of different AION scenarios to quadratic scalar DM interactions with electrons (left) and photons (right). The grey regions show parameter space that has already been excluded through searches for violations of the equivalence principle~\cite{Wagner:2012ui}, atomic spectroscopy~\cite{Hees:2016gop} or by the MICROSCOPE experiment~\cite{Berge:2017ovy}.}
\label{DMplotquad}
\end{figure}

Fig.~\ref{DMplot} shows the calculated sensitivities of AION for three such scalar DM scenarios, namely light scalar DM that couples linearly to electrons (left), to photons (middle), and through the Higgs portal 
(right), corresponding to the first, second and third terms in Eq.~\eqref{linear}, respectively.
The coloured contours show the couplings that may be detected at $\mathrm{SNR}=1$ after an integration time of $10^8$\!~s. In each case, the solid line shows the sensitivity above 0.3~Hz, while the dotted line shows the sensitivity that could be gained by extending the frequency range down to 0.1~Hz. 
We assume that the sensitivity is limited by the phase noise parameter $\delta\phi_{\rm noise}$ listed in Table~\ref{tab:parameters}. Following Ref.~\cite{Arvanitaki:2016fyj}, we have used the approximation $|\sin(x)|=\mathrm{min}\{x,1/\sqrt{2}\}$ in Eq.~\eqref{eq:deltaomega} to indicate the power-averaged sensitivity in Figs.~\ref{DMplot} and~\ref{DMplotquad}. 
For the AEDGE space experiment, the sensitive range extends down to $10^{-4}$~Hz, where gravity gradients become more important than shot noise~\cite{Arvanitaki:2016fyj}. The grey regions of parameter space have already been excluded by the indicated experiments.

We see in Fig.~\ref{DMplot} that for a scalar mass $\sim 10^{-15}$~eV the sensitivity goal for AION-10 would already improve on the limits on a scalar DM-electron coupling set by the MICROSCOPE satellite~\cite{Hees:2018fpg} by about an order
of magnitude. We also see that the initial sensitivity of AION-100 would probe an additional new range of the linear electron coupling for a scalar DM mass  $\gtrsim 10^{-15}$~eV, and begin to explore a new coupling range for the scalar-photon coupling. The sensitivity goal for AION-100 reaches deep into unexplored ranges of the linear photon and Higgs portal couplings, as does AION-km, for ultra-light scalar DM masses in the range $10^{-15}$ eV to $10^{-12}$~eV. The sensitivities of these AION variants extend far beyond the ranges currently explored by experiments with atomic clocks (see e.g.~\cite{Wcislo:2018ojh}). Finally, we note also that the space experiment AEDGE~\cite{Bertoldi:2019tck} would further extend the sensitivity to significantly lower values of the scalar DM mass and much smaller values of the linear electron, photon and Higgs portal couplings.

As seen in Fig.~\ref{DMplotquad}, AION can also explore new ranges of parameter space in models with {\it quadratic} couplings of the ultra-light scalar DM to Standard Model fields~\cite{Stadnik:2014tta}:
\begin{equation}\label{quadratic}
\mathcal{L}^{\rm{quad}}_{\rm int} =  - [\phi(\mathbf{x},t)]^2 \cdot 4 \pi G_{\rm{N}}  \times  \left[ d^{(2)}_{m_e}
m_e \overline{e}e \; - \; \frac{1}{4 }d^{(2)}_{\rm e} F_{\mu \nu} F^{\mu \nu} \right] \, .
\end{equation}
Existing limits and AION sensitivities to the quadratic coupling $d^{(2)}_{me}$ of ultra-light scalar DM to electrons are shown in the left panel of Fig.~\ref{DMplotquad}, and those for a quadratic coupling $d^{(2)}_e$ to photons are shown in the right panel.~\footnote{In  addition to the constraints displayed in Fig.~\ref{DMplotquad}, Big Bang Nucleosynthesis has the potential to impose important constraints on quadratically-coupled DM~\cite{Stadnik:2014tta}, which warrant detailed evaluation.}
We see that AION-100 will probe extensive new regions of parameter space for these quadratic couplings to electrons and photons. One feature visible in Fig.~\ref{DMplotquad} is that, if the quadratic couplings are positive, they may be screened in terrestrial experiments~\cite{Hees:2018fpg}, reducing the experimental sensitivity. This effect causes the steep rises in the atomic clock constraints and the AION sensitivity at larger masses in Fig.~\ref{DMplotquad}. On the other hand, a space-based experiment such as AEDGE is less affected by this screening, so that it maintains sensitivity at larger masses, as also seen in Fig.~\ref{DMplotquad}. The possible effect of gravitational gradient noise, which is particularly relevant for AION-km, is not included in Fig.~\ref{DMplotquad}, as it may be fully characterised and thus subtracted. 

We note that, in addition to the above, AION could be sensitive to ultra-light scalar DM via the indirect effects of the inertial and gravitational implications of the variations of the atomic masses and the mass of the Earth, as discussed in~\cite{Geraci:2016fva}. These new effects may increase the sensitivity of AION by several additional orders of magnitude of the DM couplings in the mass range of $10^{-23}$~eV to $10^{-16}$~eV, but this is still to be investigated. 

\subsection{Vector dark matter and axion-like particles}

Operating in different modes, atom interferometers such as AION can also be used to search for other ultra-light DM candidates.

\begin{figure}
\centering
\vspace{-0.5cm}
\includegraphics[width=0.6\textwidth]{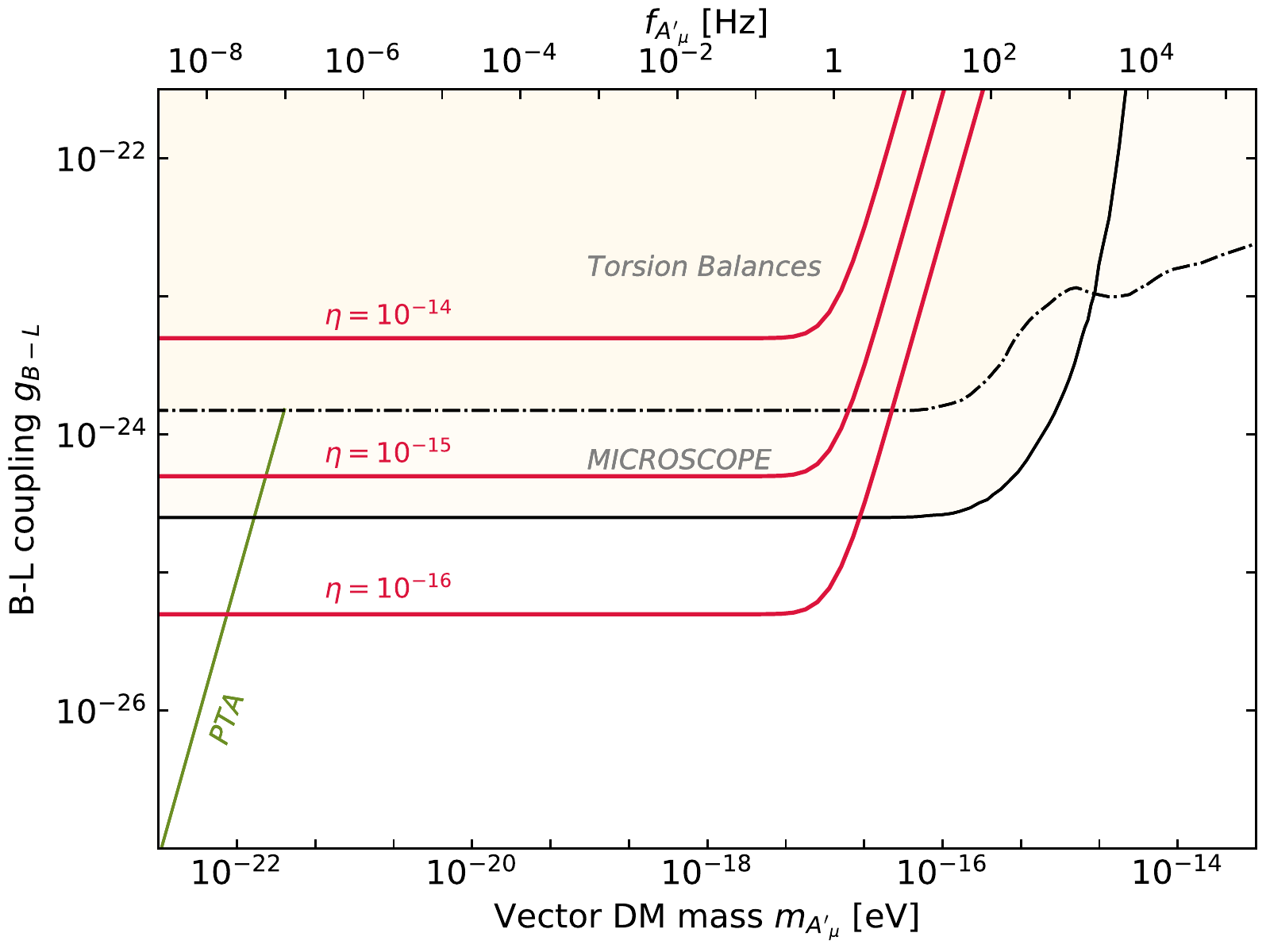}
\caption{\it The red lines show the projected sensitivities of AION to the mass and coupling of a light vector boson coupling to $B-L$, expressed in terms of the sensitivity to the E{\"o}tv{\"o}s parameter~$\eta$. We assume that $^{88}$Sr and $^{87}$Sr are used and follow the analysis in~\cite{Graham:2015ifn}. The yellow shaded region is excluded by static equivalence principle tests, and the green line indicates the pulsar timing array (PTA) sensitivity~\cite{vanHaasteren:2011ni}.}
\label{fig:vector}
\end{figure}

Using two different atomic species simultaneously, interferometers can act as an accelerometer.
In addition to allowing for a precise test of the weak equivalence principle (as discussed in section~\ref{FP}), this also gives sensitivity to, for example, a dark $B-L$ vector boson with mass $m_\phi \lesssim 10^{-15}$~eV. The AION sensitivity to this DM candidate is shown in Fig.~\ref{fig:vector}, where we assume that $^{88}$Sr and $^{87}$Sr are used and we follow the analysis in~\cite{Graham:2015ifn}. 

AION’s sensitivity to this DM candidate will ultimately be determined by the extent to which the effect of static gravity gradients can be mitigated. Indeed, static gravity gradients pose a major challenge for this kind of measurement because they give rise to spurious phase-shift contributions that couple to the initial position and velocity of the atomic wave packet and mimic weak equivalence principle (WEP) violations. This implies rather stringent requirements on how well the initial position and velocity of the two atomic clouds need to be controlled and can also lead to a significant loss of contrast~\cite{Roura:2015xsa}. Fortunately, a valuable technique for compensating the effect of gravity gradients and substantially relaxing such demanding requirements has recently been proposed~\cite{Roura:2015xsa}. The technique has been experimentally demonstrated for both gradiometry measurements~\cite{PhysRevLett.119.253201} and WEP tests~\cite{Overstreet:2017gdp} and shown to be very effective.
To gauge the sensitivity of AION, in Fig.~\ref{fig:vector}, we show projections in terms of AION's sensitivity to the $B - L$ coupling $g_{B-L}$ and the dimensionless E{\"o}tv{\"o}s-parameter~$\eta = 2 (g_A - g_B)/(g_A + g_B)$, where $g_{A,B}$ are accelerations induced by gravity for the two atomic species $A$ and $B$~\cite{PhysRevLett.119.253201}. 
The lines show the sensitivities to $g_{B-L}$ if $\eta=10^{-14}, 10^{-15}$ or $10^{-16}$ is reached~\cite{Overstreet:2017gdp,Rudolph:2019vcv}, as may be achievable with AION within an integration time of $10^8$~s. The shaded region in Fig.~\ref{fig:vector} shows the parameter space already excluded by torsion balances and MICROSCOPE~\cite{Berge:2017ovy}.

Axion-like DM particles~\cite{Marsh:2017hbv} with pseudoscalar couplings can cause nuclear spins to precess. Using atoms with different nuclear spins, atom interferometers are sensitive to such pseudoscalar nucleon couplings for axion-like particles with masses $ < 10^{-14}$~eV~~\cite{Graham:2017ivz, Alonso:2018dxy,Bloch:2019lcy}.

\subsection{Identifying a DM signal}
Confirming that the origin of a positive detection is due to DM may be challenging. Nevertheless, a number of characteristic features of DM may allow its signal to be distinguishable from that of other sources. 
For example, the frequency of the DM signal is largely set by the rest-mass $m_{\phi}$ of the DM and will remain constant up to small kinetic corrections in the angular frequency of $\mathcal{O}{(m_{\phi} |\mathbf{v}_{\phi}|^2)}$.
This is to be contrasted with the time dependence of GW signals from in-falling binary systems, where the frequency changes as the binary system evolves. 

The velocity distribution of DM will also be imprinted in the power spectral density (PSD) of the detector response~\cite{Derevianko:2016vpm, Foster:2017hbq}. Assuming the Maxwellian DM velocity distribution of the Standard Halo Model, the resulting PSD for frequencies $f_{\phi} \gtrsim m_{\phi}$ is expected to be strongly asymmetric, due to the parabolic dispersion relation for massive non-relativistic bosons. Unlike WIMP detectors, AION is expected to be particularly sensitive to DM substructures in the solar neighbourhood of the Milky Way's stellar halo, such as cold dark matter streams (see e.g.~\cite{Roberts:2018agv, Necib:2019zbk, OHare:2019qxc}). There are also annual modulations
in the DM signal (such as an ${\cal O}(7)\%$ change in signal bandwidth), and in the vector case the direction of polarisation of the DM field should be constant during the coherence time~$\tau_c$~\cite{Chaudhuri:2018rqn}.

Synergies between AION and other atom interferometers
(e.g.~AION-100 and MAGIS-100) are also expected to enable better probes of ULDM by exploiting the spatial phase information carried by the DM. The spatial dependence of the two-point scalar DM field correlation function varies as a function of the distance between two precision experiments and the angle between this spatial separation and the galactic velocity vector~\cite{Derevianko:2016vpm}. Thus, the Earth's rotation imprints a characteristic daily modulation on the combined DM signal. 
Additionally, a network of $N$ interferometers within the coherence length of the DM wave is expected to improve the sensitivity of a single device by a factor of $\sqrt{N}$~\cite{Derevianko:2016vpm}.

\section{Gravitational Waves}
\label{GW}

The LIGO/Virgo detectors found the first direct evidence for GWs via emissions
from the mergers of black holes (BHs) and of neutron stars~\cite{LIGOScientific:2018mvr}. New vistas in the
exploration of astrophysics, cosmology and fundamental physics have been opened up by these discoveries, and additional GW experiments are now in preparation
and being proposed, including upgrades of LIGO~\cite{TheLIGOScientific:2014jea} and Virgo~\cite{TheVirgo:2014hva}, the planned KAGRA~\cite{Somiya:2011np} and INDIGO~\cite{Unnikrishnan:2013qwa} experiments, and the proposed Einstein Telescope (ET)~\cite{Punturo:2010zz,Sathyaprakash:2012jk}
and Cosmic Explorer (CE)~\cite{Reitze:2019iox} experiments. Operating in a similar frequency range $\gtrsim$ few Hz, these experiments will improve on the current sensitivities of LIGO and Virgo. On a longer time-scale (planned for $\ge 2034$) LISA~\cite{Audley:2017drz} will be sensitive at lower frequencies $\lesssim 10^{-2}$~Hz. 
In addition, pulsar timing arrays (PTAs) provide sensitivity to GWs in much lower band of frequencies $\lesssim 10^{-7}$~Hz~\cite{vanHaasteren:2011ni}.

In addition to MAGIS~\cite{Graham:2017pmn}, several terrestrial cold atom interferometer experiments, such as MIGA~\cite{Canuel:2017rrp}, ZAIGA~\cite{Zhan:2019quq}, and ELGAR~\cite{Canuel:2019abg}, are currently 
in preparation or being proposed. They will
make measurements in the mid-frequency range between $10^{-2}$~Hz and a few Hz, and hence complement the LIGO/Virgo/KAGRA/ INDIGO/ET/CE and LISA detectors.

In this Section we discuss the capabilities of AION for exploring
GWs in this frequency range, using for illustration several
examples of astrophysical and cosmological sources of GWs.

\subsection{Astrophysical Sources}

Many galaxies are known to contain super-massive black holes 
(SMBHs) with masses between $10^6$ and billions
of solar masses. The Event Horizon Telescope (EHT) has recently released a first radio image of the SMBH in M87~\cite{Akiyama:2019cqa}, 
and is expected to release shortly observations of the Sgr A* SMBH at the centre of our galaxy. The LISA frequency range is ideal
for observations of mergers involving SMBHs.

However, it is not known how these SMBHs formed, and a few possibilities are discussed in~\cite{Woods:2019rqr}. Some SMBHs are known to have formed at redshifts $z \gtrsim 7$, though most SMBH growth is thought to have occurred when $z \sim 1 - 3$. It is expected that intermediate-mass black holes (IMBHs) with masses in the range 100 to $10^5$ solar masses
must exist~\cite{Mezcua:2017npy}, and there is some observational evidence for this.  As discussed in~\cite{Woods:2019rqr},
they may well have played key roles in
the assembly of SMBHs. For example, they may have formed the seeds for the accretion of material forming high-$z$ SMBHs, and/or they might have been important building-blocks in subsequent mergers forming SMBHs. 
Detecting mergers involving IMBHs over a range of redshifts may reveal how SMBHs evolved~\cite{Woods:2019rqr}.

\begin{figure}
\centering
\includegraphics[width=0.8\textwidth]{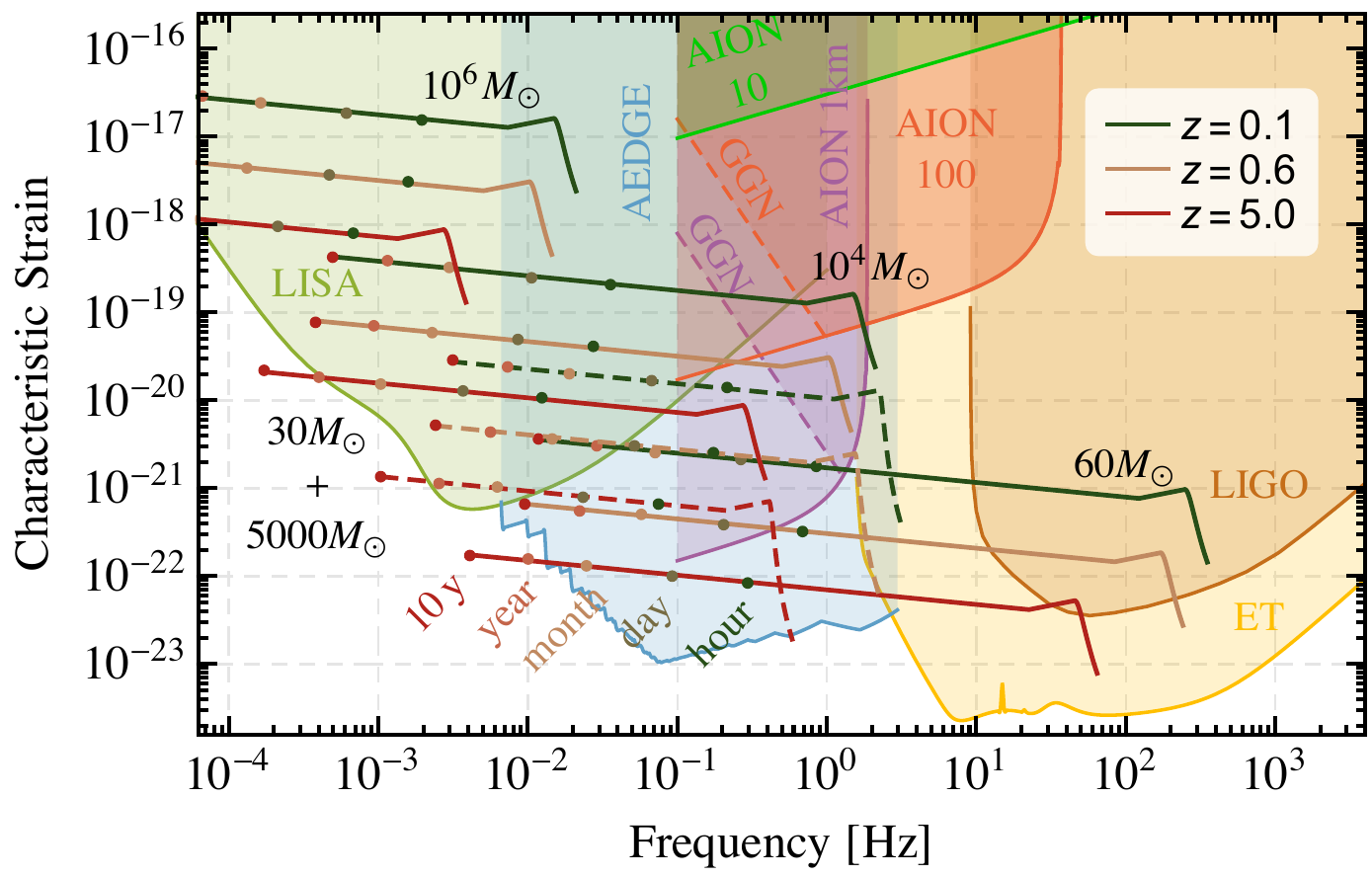}
\caption{\it Comparison of the strain measurements possible with AION and other experiments, showing their sensitivities to mergers at various redshifts of BHs of identical masses adding up to the indicated total masses (solid lines), and of an unequal-mass binary (dashed line), with the indicated remaining lifetimes before merger.  Also shown are the possible gravitational gradient noise (GGN) levels in ground-based detectors.  }
\label{staroplot}
\end{figure}

As seen in Fig.~\ref{staroplot}, the AION frequency range is ideal for observations of mergers involving IMBHs~\cite{2019arXiv191109678G}, 
which would be complementary to measurements with LISA and the LIGO, Virgo, KAGRA, INDIGO, ET or CE experiments. 
The solid lines are for the mergers of BHs with {identical masses adding up to the indicated total masses}, 
and the dashed lines are for the mergers of a stellar-class $30M_\odot$ BH with a $5000 M_\odot$ IMBH, 
all at the indicated redshifts $z$.
This figure also shows the possible gravitational gradient noise (GGN) level for 100m and 1km detectors, which may be significantly mitigated, possibly even fully measured and thus subtracted. Similar GGN levels apply to the other GW topics discussed below.

\begin{figure}[t!]
\centering
\includegraphics[height=0.42\textwidth]{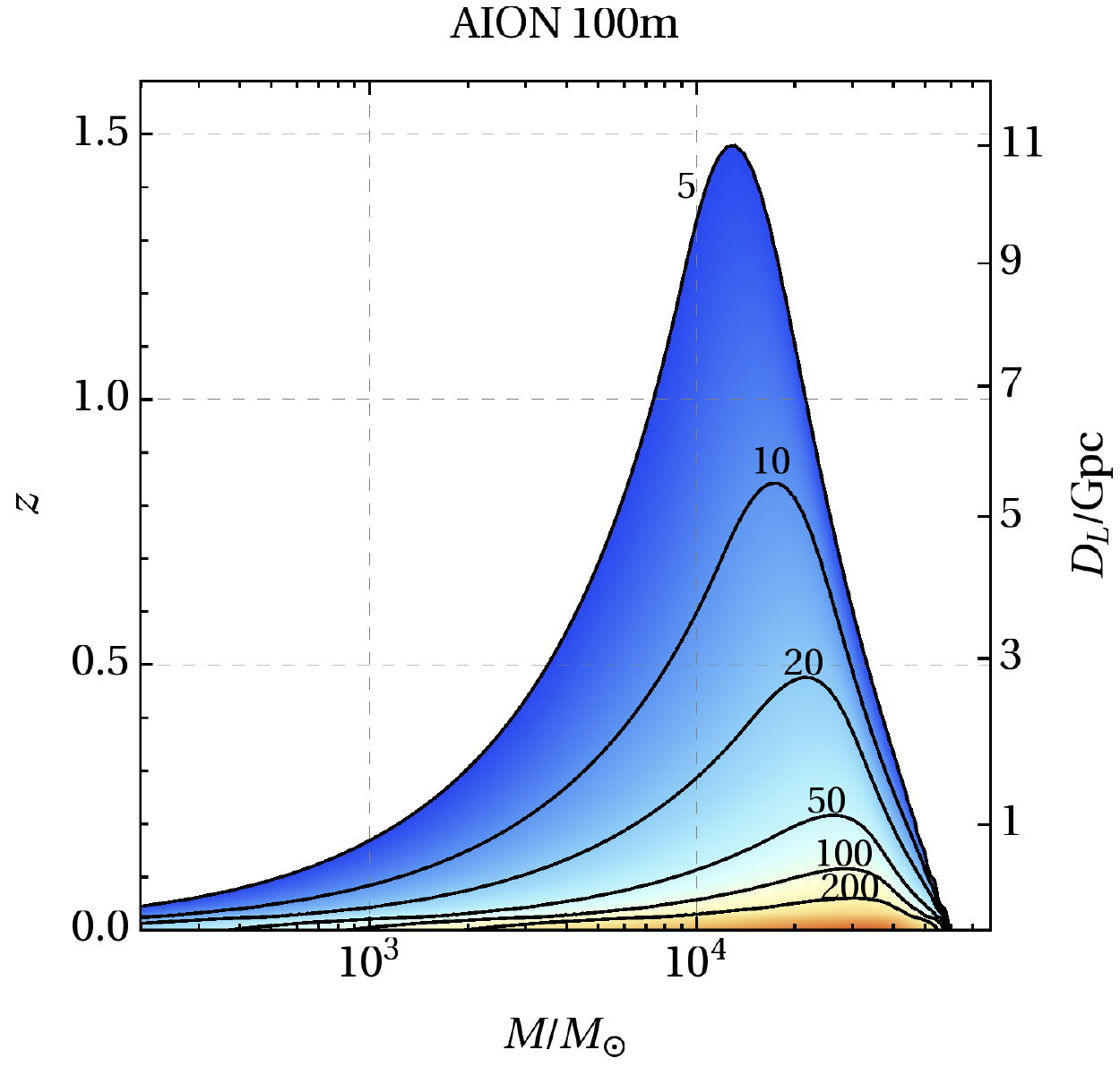} \hspace{8mm}
\includegraphics[height=0.42\textwidth]{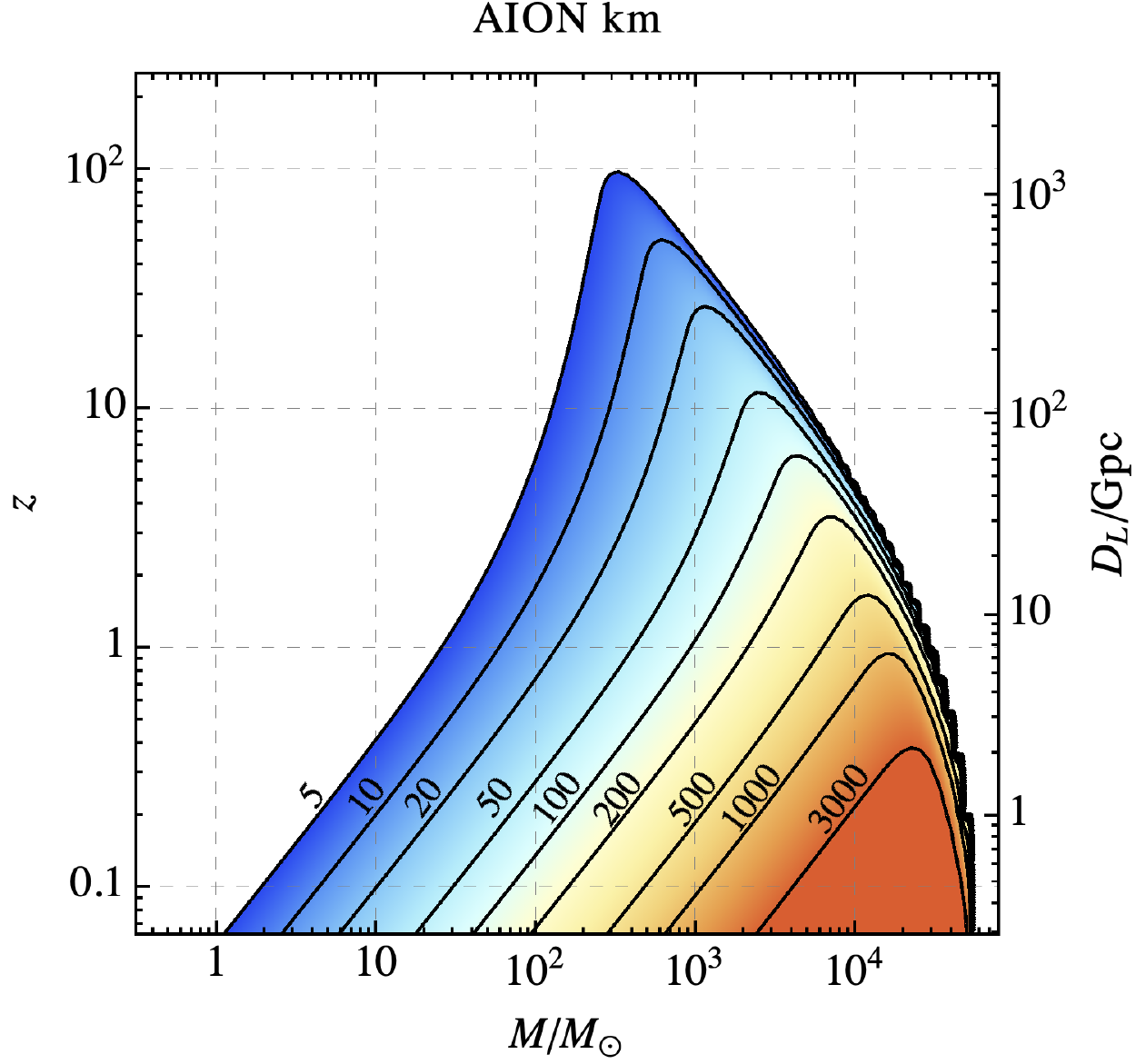}
\caption{\it Left panel: Sensitivity of AION-100 to the mergers of IMBHs with the contours showing the signal-to-noise ratio (SNR). Right panel: Similar plot for AION-1km.}
\label{staroplot2}
\end{figure}

\begin{figure}[t!]
\centering
\includegraphics[width=0.5\textwidth]{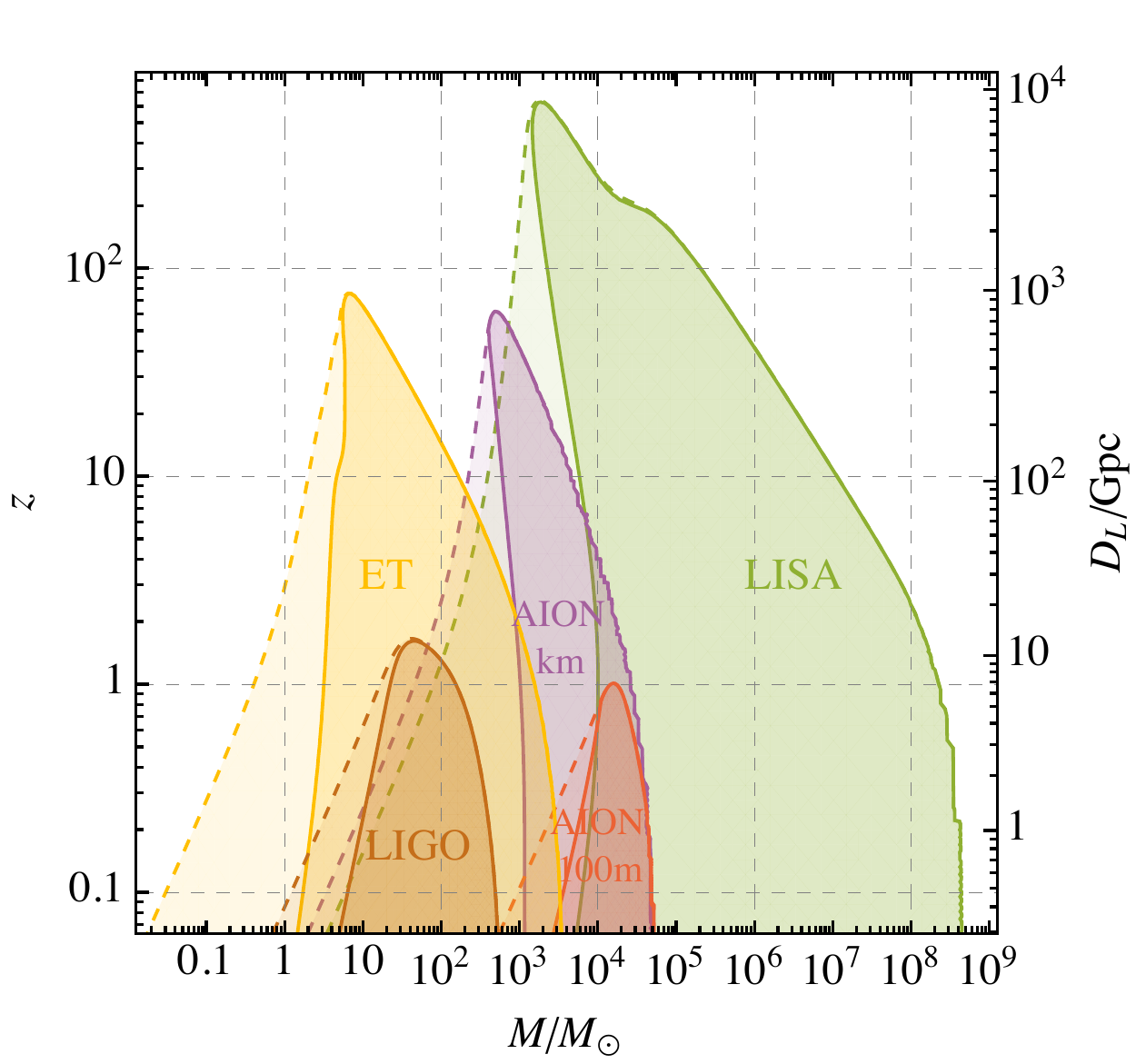}
\caption{\it Comparison of the sensitivities
of AION and other experiments with threshold ${\rm SNR}=8$.}
\label{staroplot22}
\end{figure}

 For the SNRs shown in Figs.~\ref{staroplot2} and \ref{staroplot22} we use the optimistic noise curves with fully subtracted GGN. The left panel of Fig.~\ref{staroplot2} shows the sensitivity of AION-100 for detecting GWs from the mergers of IMBHs at SNR levels $\ge 5$, which extends to redshifts $z \lesssim 1.5$ for BHs with masses $\sim 10^4$ solar masses, where there may be ${\cal O}(1)$ merger per year of BHs with such masses~\cite{Erickcek:2006xc}. The right panel of Fig.~\ref{staroplot2} shows the corresponding SNR\,$\ge 5$ sensitivity of AION-1km. In addition to the enormous SNRs attainable for the mergers of $10^4$ solar-mass black holes for $z \lesssim 1$, we see that AION-1km would be sensitive at the SNR = 5 level to mergers of $\sim 300$ solar-mass black holes out to $z \sim 100$. AION-1km would therefore be able to map out the entire assembly history of SMBHs.

 AION-1km would also be able to look for evidence of BHs with masses around $200 M_{\odot}$. We recall that low-metallicity stars with masses around this value~\cite{Heger:2002by} are expected to be blown apart by electron-positron pair-instability. The AION-1km frequency range is suitable for measuring the inspirals of BHs with masses $\sim 200 M_{\odot}$ prior to their mergers out to redshifts ${\cal O}(1)$. If such BHs are observed, they might be primordial, or have originated from higher-metallicity progenitors that are not of Population III, or have been formed by prior mergers.

Fig.~\ref{staroplot22} compares the sensitivities of various stages of the AION programme at the SNR = 8 level with those of LIGO and projections for LIGO and ET. AION-100 could complement usefully the sensitivity of LIGO in the short term, and AION-1km could complement ET in the medium term and LISA in the longer term.
Also, AEDGE~\cite{Bertoldi:2019tck} would further complement ET and LISA on a later time scale. As discussed in more detail below, simultaneous networking measurements by AION and MAGIS would refine the GW interpretations of events and refine estimates of their directions, as done currently by LIGO and Virgo, and networking AION with LIGO, ET and LISA would also offer synergies.

Finally, we note that there also is a potential ``monochromatic" astrophysical
GW signal from quantum transitions in superradiantly-produced scalar or vector clouds around BHs~\cite{Arvanitaki:2009fg,Arvanitaki:2010sy,Arvanitaki:2016qwi,Baryakhtar:2017ngi,Siemonsen:2019ebd}. For IMBHs in the Milky Way with masses in the range of a few $\times 10^3$ to $\sim 10^6$ solar masses these transitions would give frequencies inside the AION sensitivity region. 

 \begin{figure}[t!]
\centering
\includegraphics[height=0.42\textwidth]{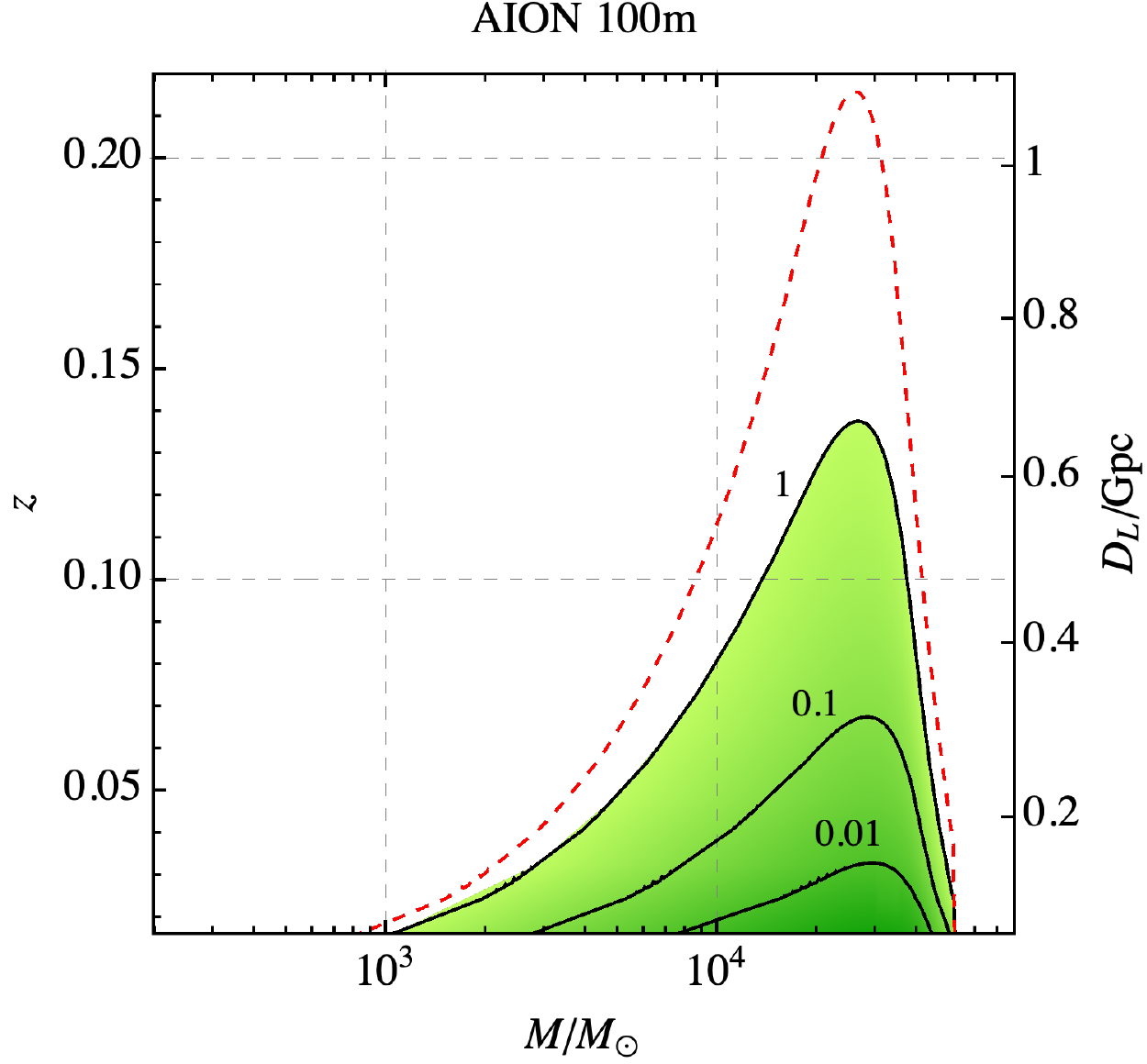} \hspace{8mm}
\includegraphics[height=0.42\textwidth]{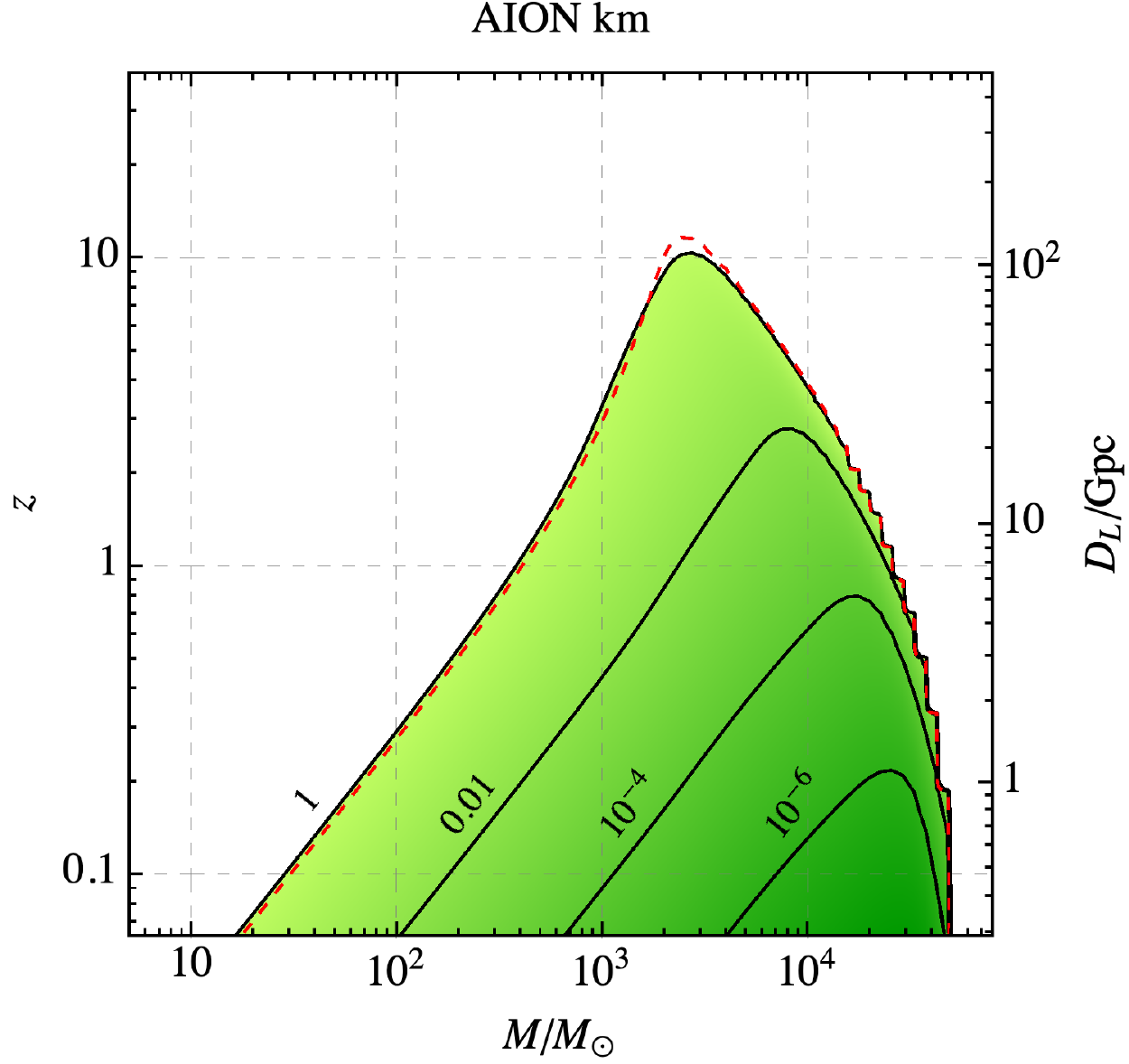}
\caption{\it Left panel: Angular pointing accuracy of AION-100 networked with MAGIS-100 with the shadings corresponding to solid angles $< (1, 10^{-2}, 10^{-4}, 10^{-6}) \times 4 \pi$. The dashed red contour corresponds to ${\rm SNR} = 50$. Right panel: Similar for AION-1km networked with MAGIS-1km.}
\label{Point}
\end{figure}

\subsection{Operating in a Network}
\label{network}

 The first networking opportunity for AION will come from operation of AION-10 in parallel with the corresponding MAGIS-10 detector at Stanford. It is planned to continue operation of AION-10 after AION-100 is commissioned, which will provide an additional networking opportunity, as well as network AION-100 with MAGIS-100. As discussed in Section~3.2, these combined operations will help in the verification of any signal for ultra-light DM.
 
 In addition to the stand-alone capabilities of AION for GW detection illustrated in Figs.~\ref{staroplot} and~\ref{staroplot2}, interesting synergies would be provided by operating AION together with MAGIS as a two-site network. For example, the pointing accuracy for two similar detectors separated by a distance $D$ observing the same signal with timing accuracies $\sigma_{1,2}$, expressed as an angular area in the sky would be
\begin{equation}
\frac{\rm {Area~(90\%~CL)}}{4 \pi} \; \simeq 3.3 \times \frac{\sqrt{\sigma_1^2 + \sigma_2^2}}{D} \; ,
\end{equation}
where $\sigma_{1,2} = 1/(2 \pi \rho_{1,2} \sigma^f_{1,2})$ with $\rho_{1,2}$ the SNRs and $\sigma^f_{1,2}$ the frequency bandwidths of the two detectors~\cite{Fairhurst:2009tc}. Assuming identical performances for MAGIS-100 and AION-100 located at Fermilab and in the UK, respectively, we find the pointing accuracies shown in the left panel of Fig.~\ref{Point}, where the shadings correspond to areas with solid-angle accuracies $< (1, 10^{-2}, 10^{-4}, 10^{-6}) \times 4 \pi$, respectively. The right panel of Fig.~\ref{Point} shows the corresponding pointing accuracies for AION-1km and MAGIS-1km, assuming for illustration that the latter is located at the Sanford Underground Research Facility. In both cases ${\rm SNR}\gtrsim 50$ is needed in order for the source to be localized significantly.~\footnote{We also envisage networking with atom interferometer experiments elsewhere, including  MIGA~\cite{Canuel:2017rrp}, ELGAR~\cite{Canuel:2019abg} and ZAIGA~\cite{Zhan:2019quq}.}

\begin{figure}[t!]
\centering
\includegraphics[width=\textwidth]{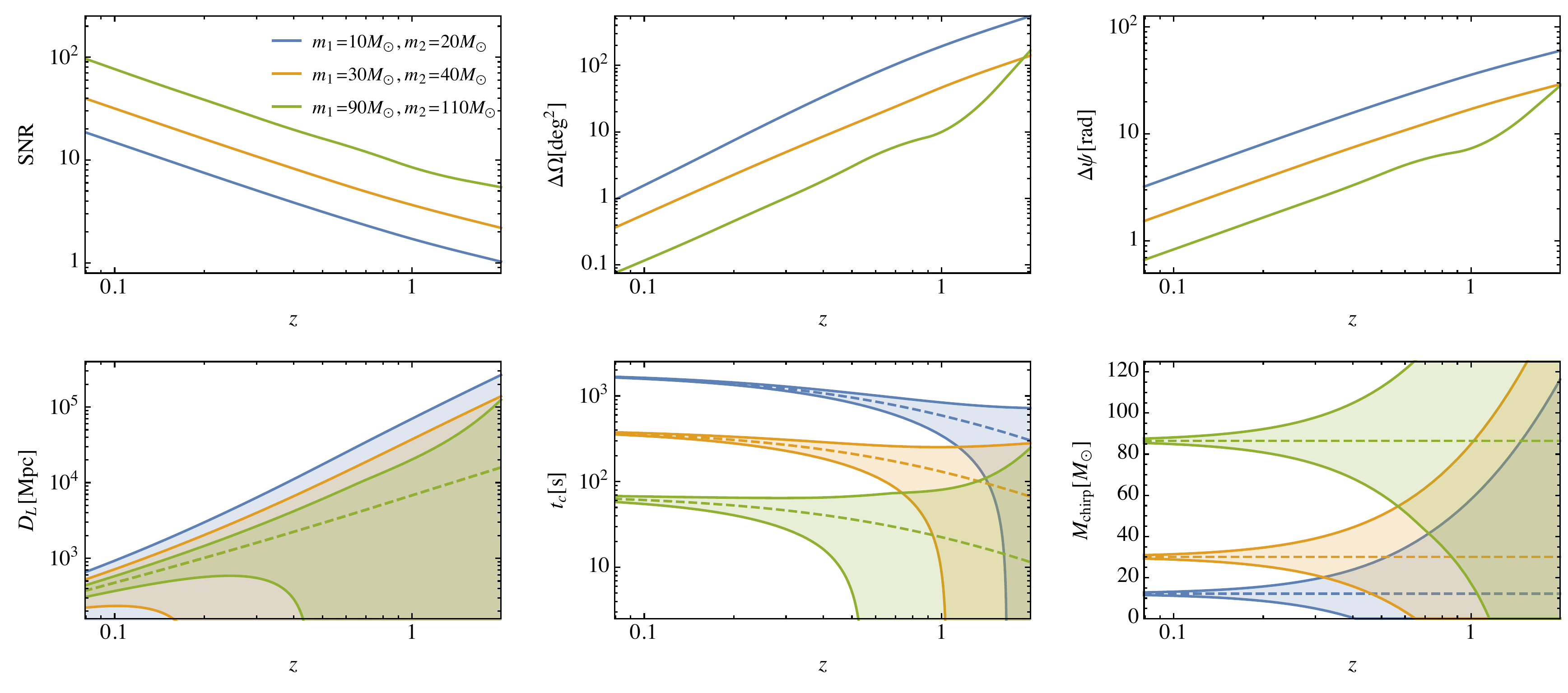}
\vspace{-6mm}
\caption{\it The SNR (upper left panel), the sky localization uncertainty $\Delta \Omega$ (upper middle panel), the polarization uncertainty $\Delta \psi$ (upper right panel), and the uncertainties in the luminosity distance $D_L$ (lower left panel), the time remaining before merger $t_c$ (lower middle panel) and the chirp mass $M_{\rm chirp}$ (lower right panel), as calculated assuming AION-km measurements for three merging binaries of different BH mass combinations as functions of their redshifts.}
\label{fig:localization}
\end{figure}

 There would also be significant synergies between AION measurements and observations in other frequency ranges. 
As seen in Fig.~\ref{staroplot}, AION could observe early inspiral stages of mergers, providing good angular localization precisions~\cite{Graham:2017lmg} and predictions for the time of the merger that could subsequently be measured by LIGO/Virgo/KAGRA/INDIGO/ET/CE. 
The inspiral phases of these sources can be observed for several months by AION,  as the detector orbits the Sun. Fig.~\ref{fig:localization} shows some examples of these possible synergies for AION-km measurements of the inspiral phases of binaries that merge in the LIGO/Virgo sensitivity window. The upper left plot shows the SNRs as functions of redshift, and the other plots show how precisely various observables can be measured by observing for 180 days before the frequency of the signal becomes higher than 3\,Hz, corresponding to the upper limit of the AION-km sensitivity window. As examples, we see in the upper middle panel that for events typical of those observed by LIGO/Virgo at $z\simeq 0.1$ the AION-km sky localization uncertainty is ${\cal O}(1)\,{\rm deg}^2$, and in the lower middle panel that the times of the mergers could be predicted with uncertainties measured in minutes, permitting advance preparation of comprehensive multi-messenger follow-up campaigns. We also see in the lower panels that for binaries at high redshifts $z\gtrsim 1$ the uncertainties in the luminosity distance, the time before merger and the chirp mass become significant
for $z = {\cal O}(1)$, though in these cases the measurements could be improved by starting to observe the binary more than 180 days before it exits the sensitivity window.

Conversely, LISA observations could be used to make predictions for subsequent AION $10^4$ solar-mass mergers. The combined measurements would also provide unparalleled lever arms for testing general relativity, measuring post-Newtonian parameters and probing Lorentz invariance in GW propagation. For example, we recall that the remnant BH mass and spin can be inferred from each of these parts separately using predictions of general relativity, and any inconsistencies between these could indicate some violation of general relativity~\cite{Ghosh:2016qgn,Ghosh:2017gfp}. In addition, the observations of overtones in the ringdown spectrum can be used to test the no-hair hypothesis of general relativity according to which mass and spin are the only properties of BHs, as has been done recently with data from GW150914~\cite{Isi:2019aib}. AION measurements would enable similar tests to be extended to much heavier BHs, as seen in Fig.~\ref{staroplot}.

Another issue that operating in a network would simplify significantly is distinguishing a stochastic background from the noise. Indeed the simplest method of overcoming this problem is running two detectors probing the same signal but with uncorrelated noise allowing for lower SNR signals to be detected~\cite{Maggiore:1999vm}. In the following section we take advantage of the network assuming the noise will be measured exactly in real time as in two uncorrelated experiments while neglecting a small modification of the sensitivity coming from the geometric configuration of the network~\cite{Thrane:2013oya}.

\subsection{Cosmological Sources}
 First-order phase transitions in the early Universe are predicted by many extensions of the Standard Model (SM) of particle physics, such as extended electroweak
sectors, effective field theories with higher-dimensional operators and hidden sector interactions. We note in particular that extended electroweak models provide
particularly interesting options for electroweak baryogenesis and magnetogenesis~\cite{Ellis:2019tjf}.~\footnote{GWs may also be produced in the very early universe by collisions of ultra-relativistic bubble walls in string scenarios~\cite{GarciaGarcia:2016xgv}.}

\begin{figure}[t!]
\centering
\includegraphics[width=0.47\textwidth]{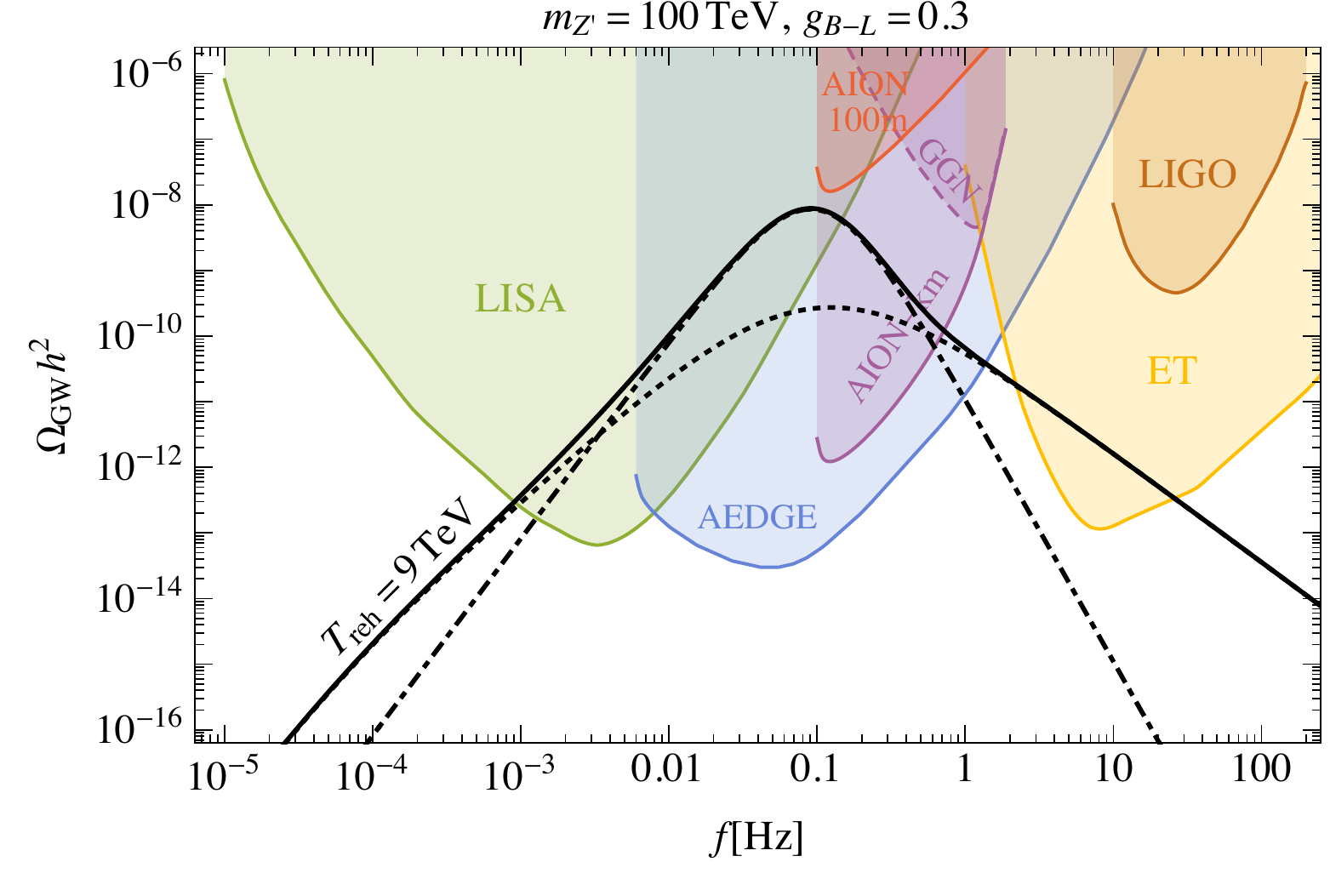}
\hspace{5mm}
\includegraphics[width=0.47\textwidth]{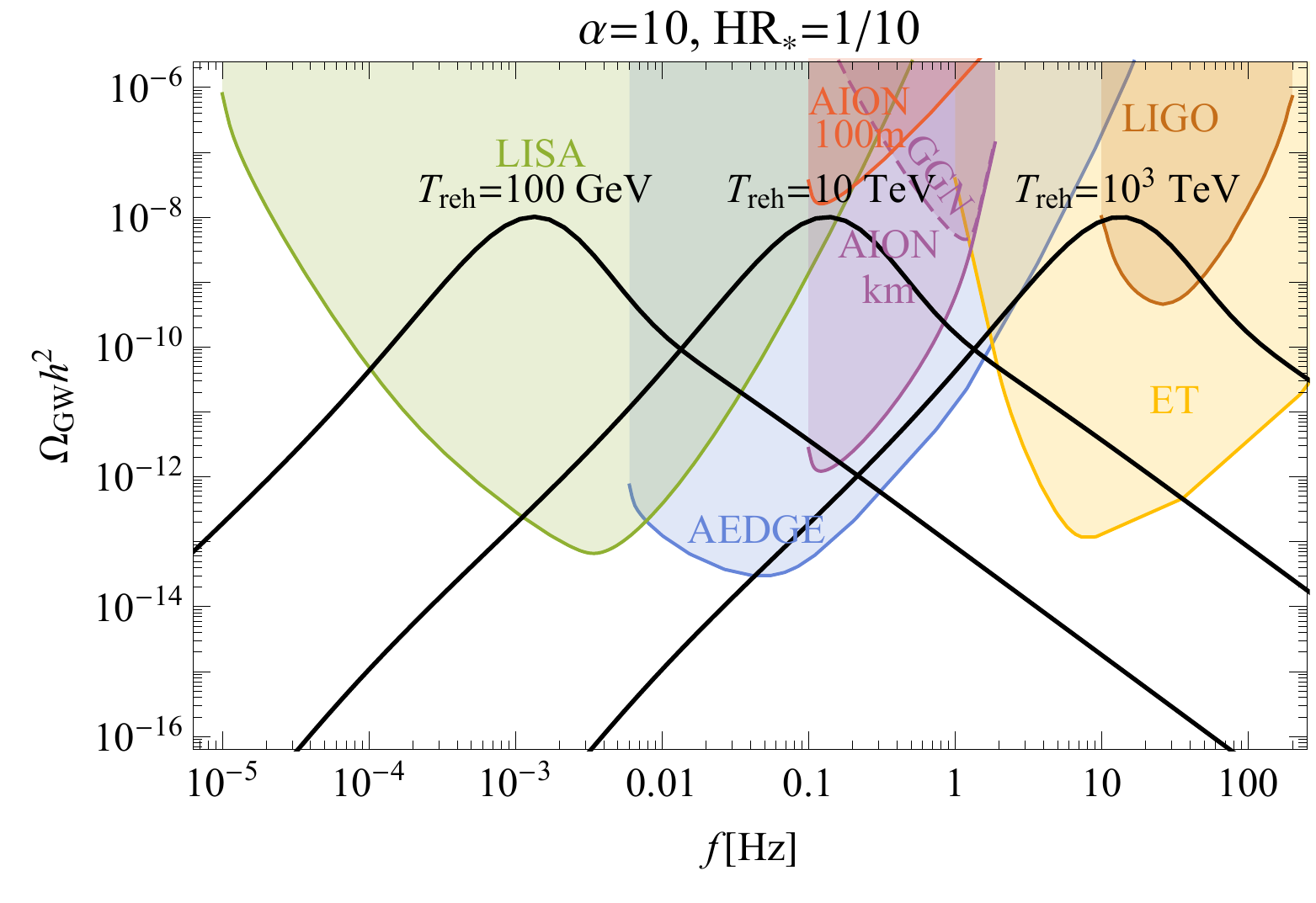}
\caption{\it Left panel: Example of the GW spectrum in a classically scale-invariant extension of the SM with a massive $Z^\prime$ boson
compared with various experimental sensitivities, including the contribution from sound waves due to bubble collisions (dash-dotted line) and turbulence in the the primordial plasma (dotted line).
Right panel: Examples of GW spectra in models of first-order phase transitions with different reheating temperatures $T_{\rm reh}$.}
\label{staroplot3}
\end{figure}

\begin{figure}[t!]
\centering
\includegraphics[height=0.38\textwidth]{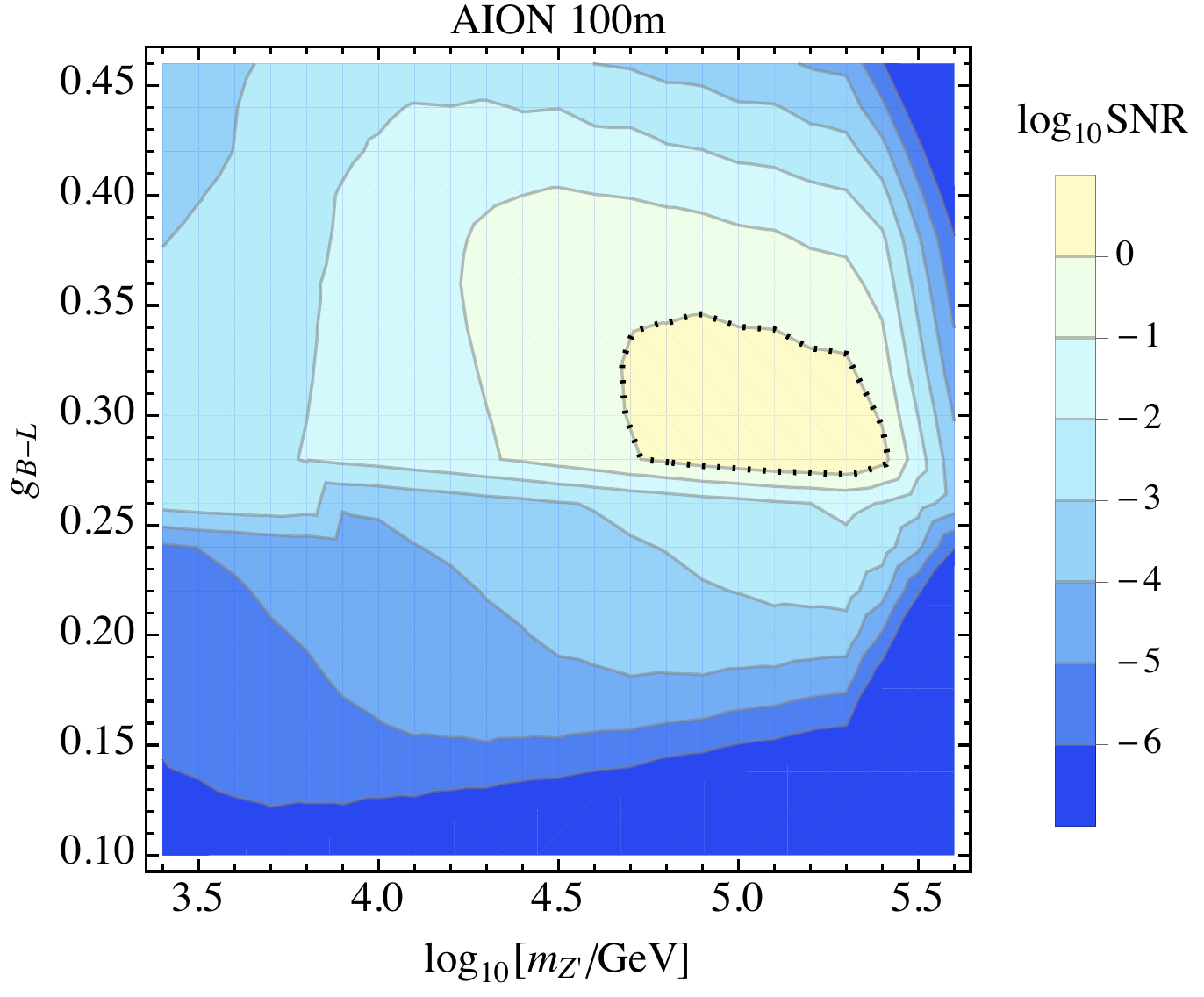}
\hspace{6mm}
\includegraphics[height=0.38\textwidth]{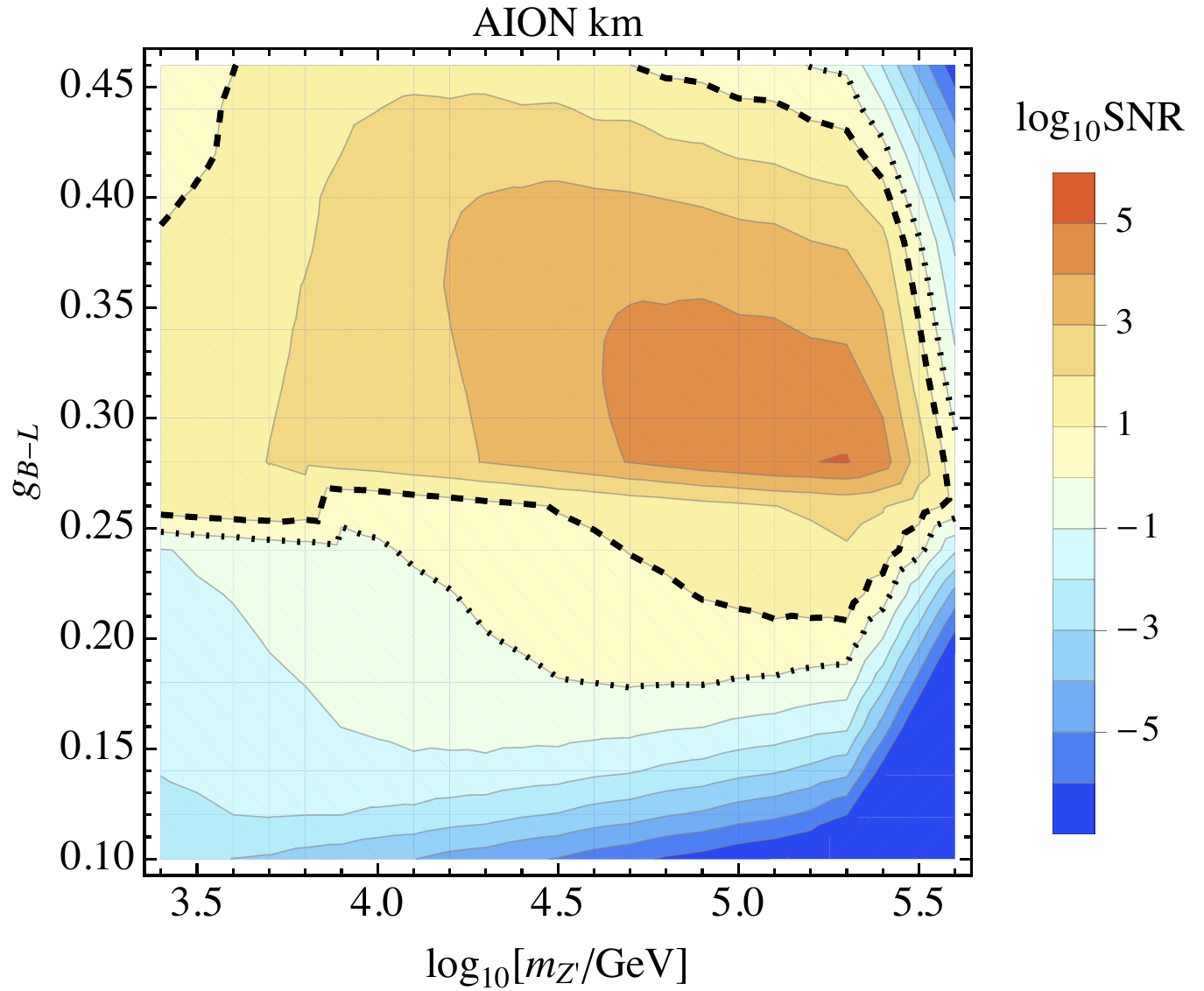}
\caption{\it SNR in the $(m_{Z^\prime}, g_{B-L})$ -plane of the scale-invariant extension of the SM for the AION-100 stage (left panel) and
the AION-1km stage (right panel). The long and short dashed lines highlight the contours ${\rm SNR}=10$ and ${\rm SNR}=1$, respectively.}
\label{staroplot32}
\end{figure}

An example of the GW spectrum calculated in a classically scale-invariant extension of the SM with a U(1)$_{B-L}$ $Z^\prime$ boson in the left panel of Fig.~\ref{staroplot3}. In the example shown, $m_{Z^\prime} = 100$~TeV and its coupling $g_{B-L} = 0.3$, which corresponds to the maximum SNR of 3.3 for AION-100 seen in the left panel of Fig.~\ref{staroplot32}. Our calculation includes GWs sourced by both sound waves  
(dash-dotted line) and turbulence in the the primordial plasma (dotted line)~\cite{Ellis:2019oqb}. These contributions yield a broad spectrum whose shape can be probed only by a combination 
of LISA and a mid-frequency experiment such as AION. In any scenario for a first-order phase transition, there is a characteristic temperature, $T_*$, at which bubbles of the new vacuum percolate to complete the transition, which depends on the model parameters and is $T_* = 220$~GeV in the
particular example displayed. Another key temperature in such a first-order transition is the temperature, $T_{reh}$, to which the universe reheats once the transitions is completed. The model illustrated in the left panel of Fig.~\ref{staroplot3} has $T_{reh} = 9.1$~TeV, and the right panel shows the total GW
spectra for a characteristic range of models with different values of $T_{reh}$. We see that AION can play a key role in distinguishing such a signal over a wide range of percolation temperatures.

\begin{figure}[t!]
\centering
\includegraphics[width=0.48\textwidth]{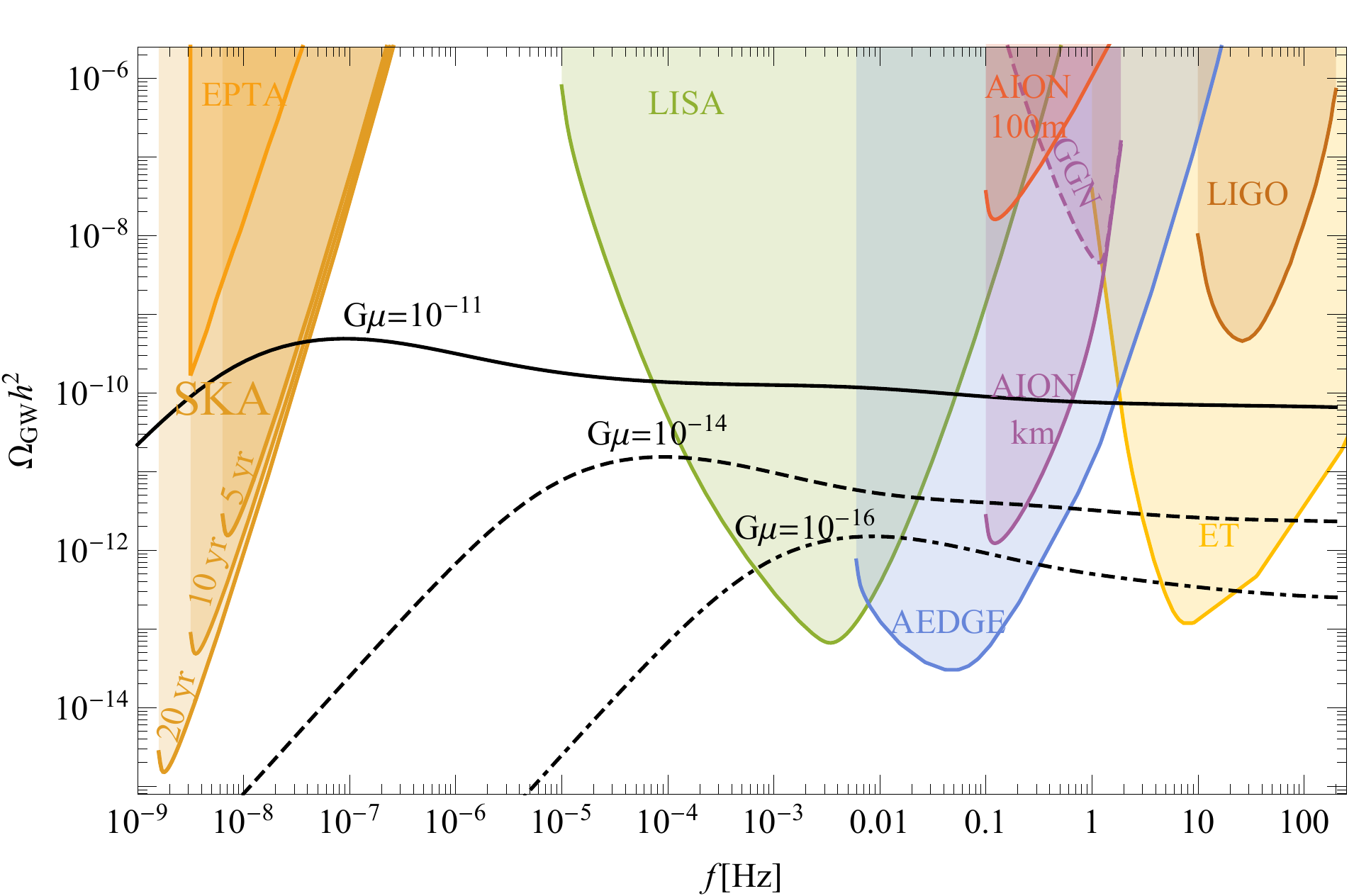}
\includegraphics[width=0.48\textwidth]{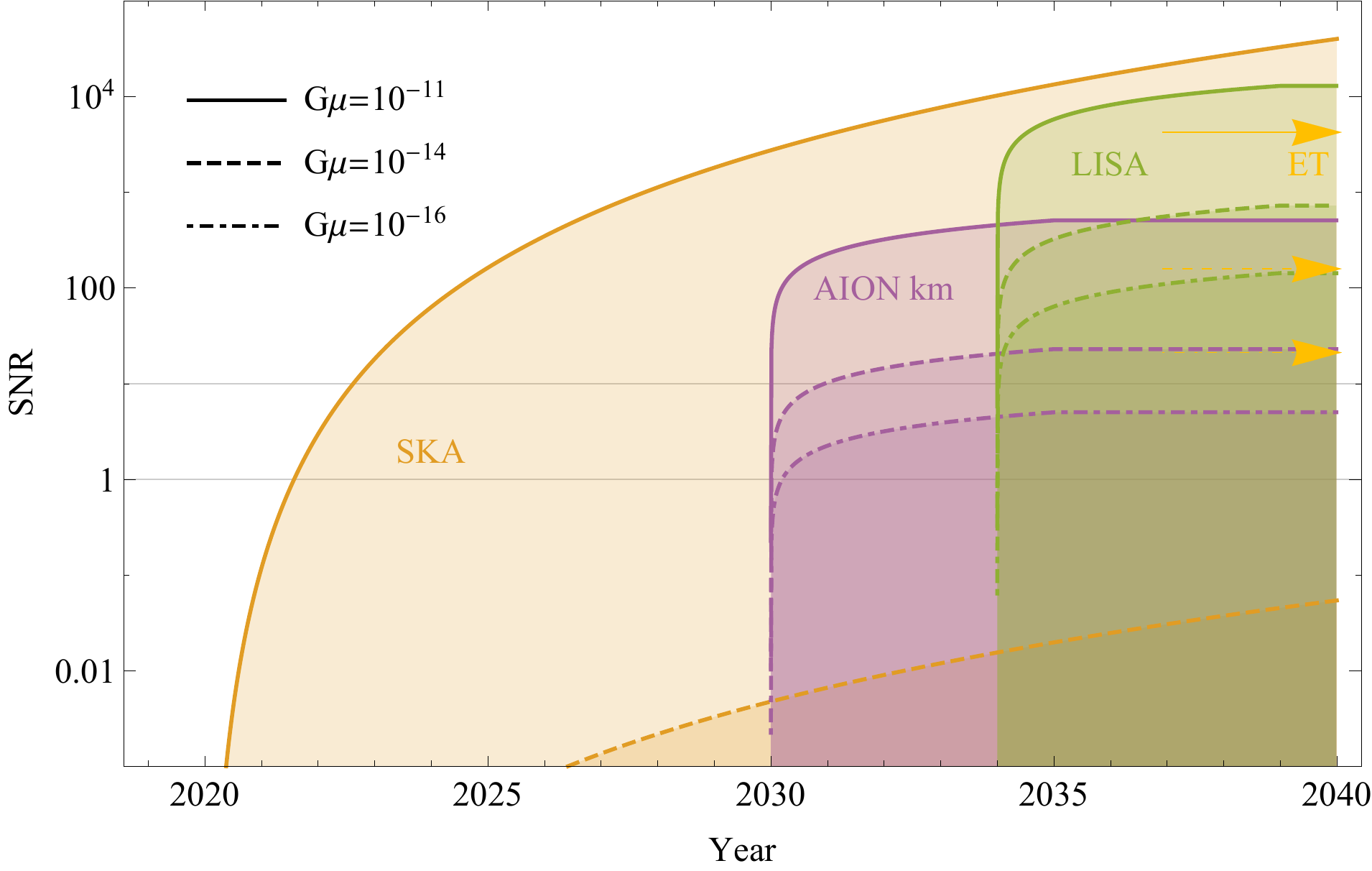}
\caption{\it Left panel: Examples of GW spectra from cosmic strings with differing tensions $G\mu$. Right panel: The SNRs for detection by AION-1km of cosmic strings with different values of $G\mu$ compared with those of the approved experiments SKA and LISA, as functions of time. Also indicated by arrows are the prospective sensitivities of ET after 5 years of operation.}
\label{staroplot4}
\end{figure}

 The discovery sensitivity of AION in the parameter space of this scale-invariant extension of the SM is indicated in Fig.~\ref{staroplot32}. Possible values of the SNR in the two-dimensional parameter space of the $Z^\prime$ boson and coupling are colour coded. We see that, while claiming some hint of detection with the 100m version might be possible (left panel), far stronger evidence would be provided by AION-1km over a much wider range of parameter space (right panel).

Cosmic strings are another possible cosmological source of GW signals. These yield a very broad frequency
spectrum stretching across the ranges to which the LIGO/ET, AION/MAGIS, LISA and SKA~\cite{Bacon:2018dui} experiments are sensitive, as seen in the left panel of Fig.~\ref{staroplot4}. Pulsar timing array (PTA) measurements at low frequencies set the current upper limit on the string
tension of $G \mu \simeq 10^{-11}$~\cite{vanHaasteren:2011ni}.

In the absence of other new physics, this limit would put the signal beyond reach of the initial AION-100 configuration, while the right panel of Fig.~\ref{staroplot4} shows that the km version would be able to discover cosmic strings with $G\mu \sim 10^{-14}$ at the 5-$\sigma$ level. This panel also shows the prospective sensitivities of SKA~\cite{Bacon:2018dui} observations of 1000 pulsars and LISA. The sensitivities of these approved projects are indicated as functions of time, according to their announced schedules. We note that SKA should be able to surpass the sensitivity of AION-1km for $G\mu = 10^{-11}$, but would not be competitive for $G\mu = 10^{-14}$ or $10^{-16}$, while in the longer term LISA would be the most sensitive for all the values of $G\mu$ studied. We also indicate by arrows the prospective sensitivities of ET after 5 years of operation.

\begin{figure}
\centering
\includegraphics[width=0.5\textwidth]{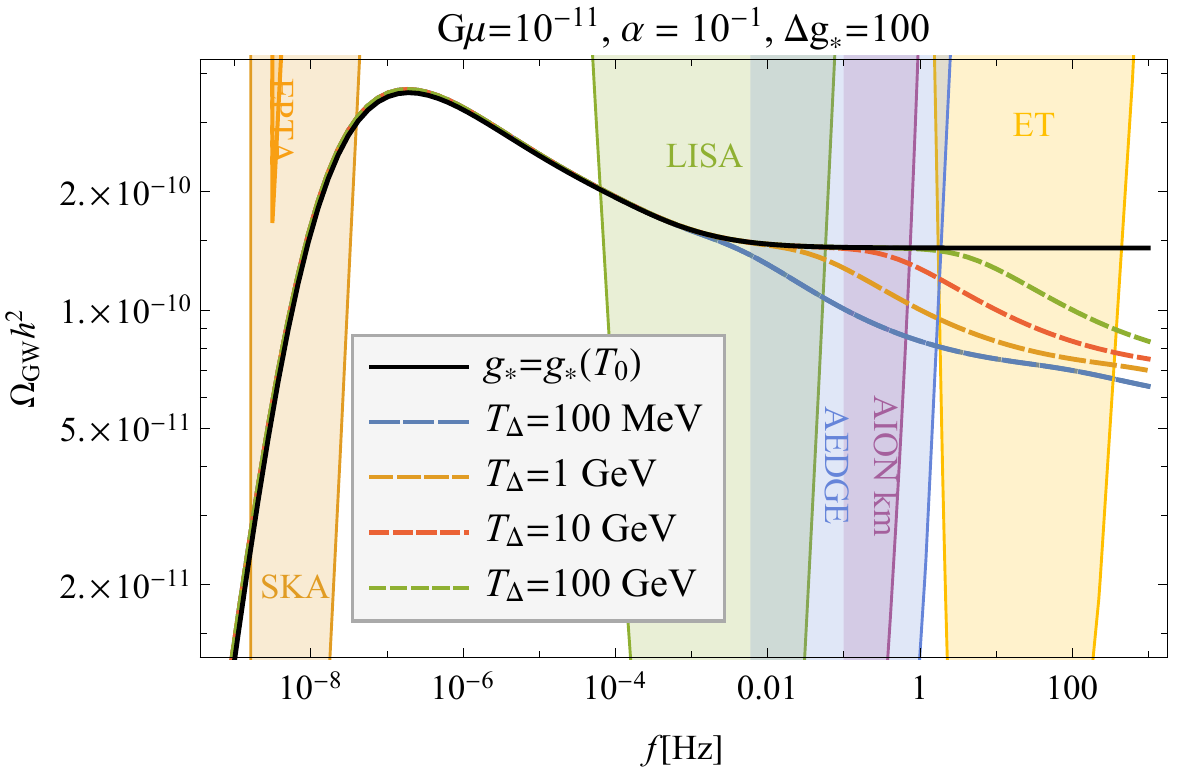}
\includegraphics[width=0.47\textwidth]{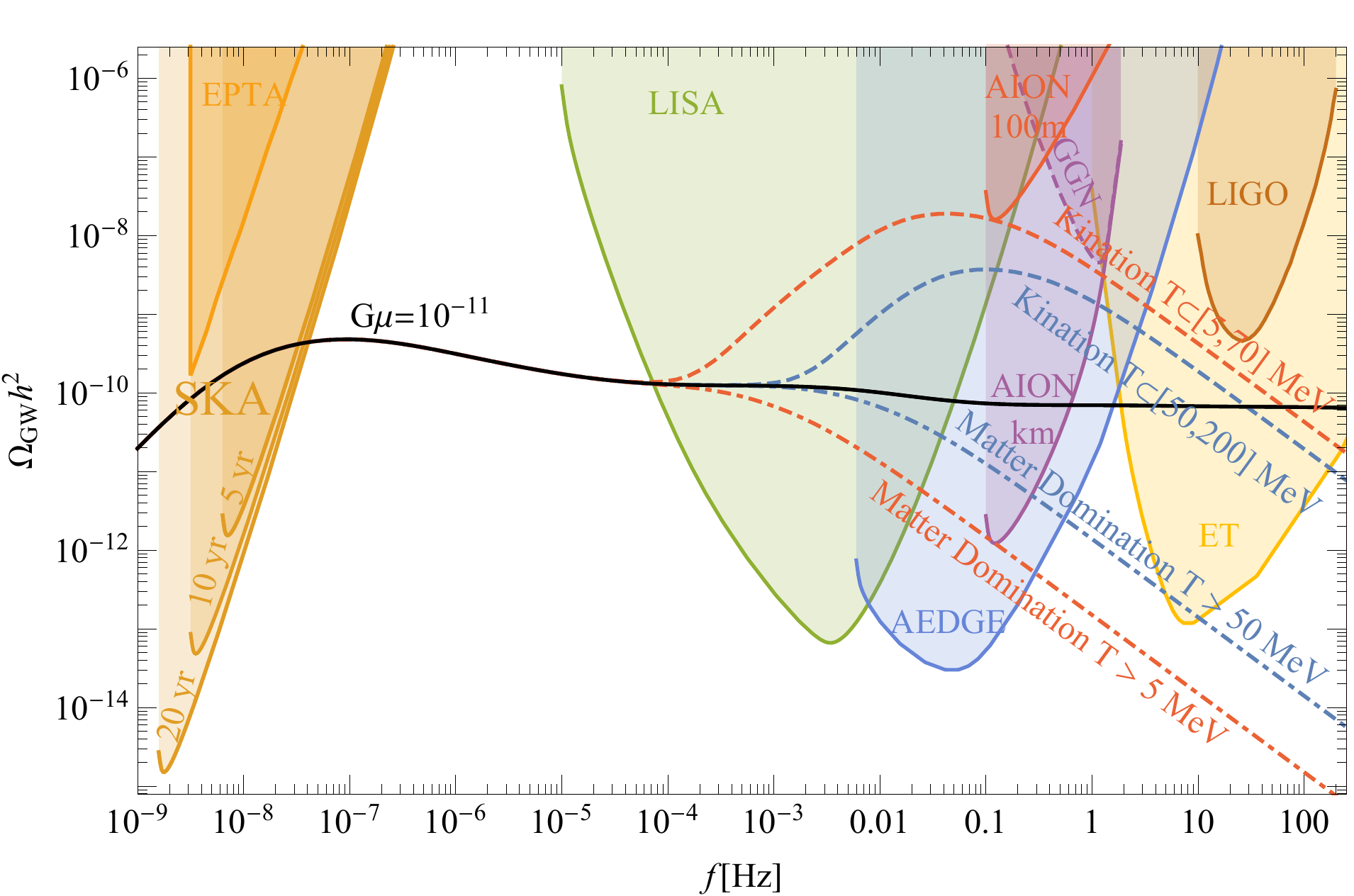}
\caption{\it Left panel: Effect on the GW spectrum for the
case $G \mu = 10^{-11}$ of a new particle threshold at various energies $T_\Delta \ge 100$\,MeV with an increase $\Delta g_* = 100$ in the number of relativistic degrees of freedom. Right panel: The effect on the GW spectrum for the case $G \mu = 10^{-11}$ of modified cosmological scenarios with episodes of of matter domination at temperatures $> 5$ and $> 50$~MeV and of kination at temperatures between 5 and 70~MeV and between 50 and 200~MeV.}
\label{staroplot5}
\end{figure}

Fig.~\ref{staroplot5} shows the possible effects of modifications of the conventional cosmic expansion history on the GW spectrum generated by cosmic strings. The left panel illustrates the impacts of changes in the number of relativistic degrees of freedom by $\Delta g_* = 100$ at various temperatures between 100~MeV and 100~GeV, as might occur in some scenarios with dark sectors. This example demonstrates that probing the expected plateau in the cosmic string GW signal over a wide range of frequencies with different detectors can provide significant information on the evolution of the universe as well as on cosmic strings themselves~\cite{Cui:2017ufi,Cui:2018rwi}. {For example, if there is a mismatch between the levels of cosmic string signals at LISA and higher-frequency experiments such as LIGO, AION measurements could be crucial for pinpoint the energy scale and nature of the mechanism responsible.} This point is also illustrated in the right panel of Fig.~\ref{staroplot5}, where we see that a period of matter domination at temperatures $T > 5$~MeV could suppress the GW signal below the reach of ET while remaining within the reach of AION-1km. On the other hand a period of kination would enhance the GW spectrum, eg., kination at temperatures $T \in [5, 70]$~MeV could bring GWs from cosmic strings within the reach of AION-100.

\section{Other Fundamental Physics}
\label{FP}

There are prospects for studying other aspects of fundamental physics beyond dark matter and GWs using ultra-high-precision atom interferometry that are still to be investigated in detail. The design of AION has not been optimised for such studies, but it may have interesting capabilities for such studies, such as:

\noindent
$\bullet$  {\it Other DM-SM interactions}. We explored in Section~\ref{ULDM} atomic interferometry signatures of ULDM due to linear interactions with the SM (see Eq.\eqref{linear}) and quadratic interactions (see Eq.~\eqref{quadratic}). Atom interferometers such as AION may also be sensitive to other couplings. For example, explorations of effective couplings along the lines of other DM experiments, see, e.g., Ref.~\cite{Fitzpatrick:2012ix}, may provide interesting new opportunities.\\
$\bullet$ {\it Probes of long-range fifth forces}:  As an example, since atom interferometry is capable of detecting the gravitational field of Earth \cite{Peters_2001}, a combination of interferometers at different heights would make it possible to explore the existence of any other long-range fifth force coupling to matter differently from gravity. Searches for such long-range forces have connections to models of dark matter and modified gravity and form a very active area of research on physics beyond the SM, see, e.g.,  \cite{Safronova:2017xyt}. Classical searches for fifth forces with universal Yukawa-type couplings on terrestrial scales are significantly constrained by ~\cite{Adelberger:2003zx,Safronova:2017xyt}. \\
$\bullet$ {\it Tests of general relativity}:  Arrangements of atom interferometers at different heights also enable measurements of higher-order general-relativistic corrections to the Earth's gravitational potential.  The leading higher-order effects include those due to the potential gradient, corrections due to the finite speed of light, and D\"oppler effects on the photon frequency~\cite{ Dimopoulos:2008hx}. 
The current direct sensitivity to the graviton mass, $m_g$, using the speed of GW propagation could be improved using AION measurements, exploiting its potential for longer-duration observations and the detection of the inspirals of higher-mass black holes whose frequencies would be lower than those measured by LIGO and Virgo~\cite{Carson:2019kkh}. For example, preliminary estimates indicate that AION-km measurements of an event resembling GW170104 would be able to set an upper limit $m_g \sim 1.7 \times 10^{-24}$~eV at the 95\% CL, compared with the upper limit of $7.7 \times 10^{-23}$~eV from GW170104, and a event involving two 100-solar-mass black holes would give AION-1km sensitivity to $m_g < 1.2 \times 10^{-24}$~eV. Also, using different atomic species, AION could probe the equivalence principle to the level of $10^{-16}$, as illustrated in Fig.~\ref{fig:vector}. \\ 
$\bullet$ {\it Test of atom neutrality}: Atom interferometers can be used to test atom neutrality~\cite{Arvanitaki:2007gj}. AION-10 should be able to probe this to 30 decimal places, far beyond the present experimental sensitivity to 22 decimal places.\\
$\bullet$ {\it Constraining possible variations in fundamental constants}: Measurements using atom interferometers at different time and space positions may be combined to probe possible variations in fundamental constants. Some motivations for such studies can be found in \cite{Uzan:2002vq}.\\
$\bullet$ {\it Probing dark energy}:  This is currently causing the expansion of the universe to accelerate. It is expected to be present everywhere, and precise experiments can in principle be used to probe its local effects. Some models of dark energy involve  ultra-light dynamical fields that would induce space- and time-dependent modifications of fundamental physical properties, similarly to the ULDM discussed in Section~\ref{ULDM}. In other models dark energy is associated with a modification of the laws of gravity, to which atom interferometry experiments are particularly sensitive, providing important constraints on popular models \cite{Sabulsky:2018jma, Jaffe:2016fsh}. If a signal of a local modification of gravity were seen by the AION experiment, further confirmation would be required before it could be said to be due to the dynamics of dark energy. This would require other measurements, much as searches for dark matter at a particle collider would require confirmation from other experiments.  That said, AION will certainly be testing the same models of dark energy as are also currently key targets for upcoming large-scale cosmological surveys;\\
$\bullet$ {\it Probes of basic physical principles}.  Atom interferometers may be used to probe the foundations of quantum mechanics and Lorentz invariance. For example, some ideas that modify the standard postulates of quantum mechanics concerning linearity and wave-function collapse may be probed via precise interferometry of quantum states, see, e.g., \cite{Ellis:1983jz,Banks:1983by,Ghirardi:1985mt,Weinberg:2016uml}. We note also that it was proposed in Ref.~\cite{Chung:2009rm} to use atom interferometers to test the principle of Lorentz invariance in gravitation.

\section{Final Remarks}
\label{FR}

AION is a proposed programme to use quantum technologies to address fundamental physics problems. Specifically, it will use cold atom interferometers to address fundamental questions such as the nature of dark matter, and will measure gravitational waves in the mid-frequency band where astrophysical processes involving intermediate-mass black holes and cosmological phenomena such as first-order phase transitions and cosmic strings can be probed.
 
 \vspace{2mm}
\noindent As we have discussed, AION is a four-stage programme:
\begin{itemize}
\item Stage  1  is to  build  and  commission  a  10m-scale atom interferometer to be operated as part of a network,  and in parallel to develop  existing  technology, plans  and infrastructure  for  a  100m  detector. This is currently the subject of a proposal to the UK STFC and EPSRC;
\item Stage 2 would build, commission and exploit the 100m detector and prepare a design study for the km-scale detector;
\item Stage 3 would build, commission and exploit a kilometre-scale terrestrial detector, and;
\item Stage 4 would build, commission and exploit a satellite-based interferometer.
\end{itemize}

 \vspace{2mm}
\noindent The following are the primary objectives of the AION science programme:
\begin{itemize}
\item To explore well-motivated ultra-light dark matter candidates with sensitivities orders of magnitude beyond current bounds, an objective to which AION-10 can make significant contributions;
\item To explore mid-frequency band gravitational waves from astrophysical sources and from the very early universe, an objective to which AION-100 can make significant contributions;
\item To operate the successive stages of AION in a network with the MAGIS detector and potentially other atom interferometers, as well as approved laser interferometers such as LIGO, Virgo, KAGRA, INDIGO and LISA, and other proposed detectors.
\end{itemize}

 \vspace{2mm}
The AION project offers valuable synergies between traditional particle physics, astrophysics and the physics of the early universe, notably in probing the nature of dark matter and using gravitational waves to probe the formation of massive black holes and phenomena in the early universe. This programme will nurture a community built upon continuous dialogue between atomic physicists, particle physicists, astrophysicists and cosmologists, whose combined expertise will be essential to resolve these fundamental scientific issues in the long run.

\section*{Acknowledgements}
We would like to thank the MAGIS Collaboration, especially Peter W.\ Graham and Jason M.\ Hogan, for helpful discussions. We also like to thank Albert Roura for input regarding the effect of static gravity gradients and means to mitigate it. The authors acknowledge support by the UK Engineering and Physical Sciences Research Council (EPSRC) and Science and Technology Facilities Council (STFC). The work of JC was supported in part by the Royal Society. The work of JE, ML and VV was supported by STFC Grant ST/P000258/1, and JE also received support from the Estonian Research Council via a Mobilitas Pluss Grant. The work of CM was supported by STFC Ernest Rutherford Fellowship, via Grant ST/N004663/1.

\bibliographystyle{JHEP}
\bibliography{AION}

\end{document}

{
{\bf Diego~Blas},
{\bf Oliver Buchmueller},
{\bf Themis Bowcock},
{\bf Jon Coleman},
{\bf John~Ellis},
{\bf Chris Foot},
{\bf Val Gibson},
{\bf Martin Haehnelt},
{\bf Marek~Lewicki},
{\bf John March-Russell},
{\bf Christopher~McCabe},
{\bf Ian Shipsey} and
{\bf Ville~Vaskonen